\documentclass[%
 reprint,
 amsmath,amssymb,
 aps,
]{revtex4-2}

\usepackage[usenames, dvipsnames]{xcolor}
\usepackage{graphicx}
\usepackage{dcolumn}
\usepackage{bm}
\usepackage{braket}
\usepackage{hyperref}
\usepackage{url}
\usepackage{comment}
\usepackage{caption}
\usepackage{subcaption}
\usepackage{mwe}

\begin{document}

\preprint{APS/123-QED}

\title{\textit{Ab initio} study of Coulomb drag driven electron-hole bifluidity in doped graphene}

\author{Dwaipayan Paul}
\email{Contact author: dwaipayan.paul@hu-berlin.de}
\author{Elena Trukhan}
\email{Contact author: elena.trukhan@physik.hu-berlin.de}
\author{Nakib H. Protik}%
\email{Contact author: nakib.protik@physik.hu-berlin.de}
\affiliation{%
 Institut f\"{u}r Physik and CSMB, Humboldt-Universit\"{a}t zu Berlin
}

\date{\today}

\begin{abstract}
Motivated by the notion that a preponderance of Coulomb interactions might lead to hydrodynamics, we carry out an \textit{ab initio} calculation of the charge carrier transport properties of the electron-hole gas of doped graphene. We include both the phonon and Coulomb interactions within a momentum and band resolved Boltzmann transport formalism. We find that under suitable conditions, the strong Coulomb drag effect induces phenomena like negative conductivity and joint electron-hole hydrodynamics (bifluidity) in the system. We also identify the exclusive electron or hole hydrodynamics. We find that there is a strong violation of the Wiedemann-Franz law in the low doped regimes. Our work elucidates the roles of the microscopic scattering mechanisms that drive the hydrodynamic phenomena.

\end{abstract}

\maketitle

\paragraph*{Introduction} The hydrodynamic transport has been a longstanding topic of curiosity in the condensed matter and the broader materials science communities. It is a low dissipation mode of transport, characterized by a large number of scattering events that, however, are (quasi)momentum conserving. In such a non-equilibrium state, a system of particles drifts together with a characteristic global velocity, akin to how water flows through a pipe. This mode of transport lies between the low temperature ballistic and the high temperature diffusive regimes, and has been rather difficult to observe and control in 3D systems, usually due to the inevitable presence of impurity interactions that tend to destroy momentum. 2D materials, on the other hand, allow a way to introduce charge carriers via electrostatic gating, i.e., without the need to introduce dopants. As such, these systems are well-suited for studying hydrodynamics.

Now, graphene \cite{geim2007rise}, since its discovery, has been a fertile ground for studying the properties of 2D materials. This is partly because it is easy and inexpensive to obtain, and for its superior charge carrier tunability. It can sustain high charge conduction states \cite{das2011electronic,castro2009electronic}. And, being a semimetal, it simultaneously hosts both an electron and a hole subsystem. Pertinent to this work is the fact that the charge carrier hydrodynamics in this material has been predicted in several experimental and theoretical studies \cite{crossno2016observation, krishna2017superballistic, sulpizio2019visualizing, muller2009graphene, bistritzer2009hydrodynamic, narozhny2017hydrodynamic, lucas2018hydrodynamics,ku2020imaging}. Previous works \cite{muller2009graphene, crossno2016observation, ku2020imaging}, established the formation of strongly interacting Dirac fluid at the charge neutrality point (CNP), which breaks the Fermi liquid picture and exhibits a strong violation of the Wiedemann-Franz law \cite{principi2015violation,crossno2016observation, tu2023wiedemann}. Adding to the complexity, there is also formation of local impurities called charge puddles \cite{samaddar2016charge, lucas2016transport}, which provides a strong scattering channel for the charge carriers at low temperatures. Our focus here is on doped graphene closer to room temperature where we can safely ignore charge puddle effects \cite{bandurin2016negative, principi2016bulk}. Furthermore, the existence of hydrodynamics was predicted to be not only in the Dirac fluid regime, but also in the Fermi liquid regime\cite{principi2016bulk}. The latter is the focus of our work.

From a theoretical point of view, it is interesting to ask what types of scattering processes dictate the charge conductivity of graphene. Interactions of the charge carriers with phonons? With the sample boundary? Interactions between the charge carriers via Coulomb interactions? Since graphene hosts both an electron and a hole gas, it is important to ask whether the alleged hydrodynamic state is only in one or the other subsystem? Or is it of a more exotic kind where both the electron and hole subsystems share a joint global drift velocity, forming what is called as an electron-hole bifluid?

So far, answers to such questions necessarily had to employ semi-empirical transport theories  \cite{narozhny2015hydrodynamics, principi2016bulk, muller2009graphene} that contain simplified assumptions and tunable parameters. In fact, current publicly available versions of state of the art \textit{ab initio} Boltzmann electronic transport equation (BTE) solvers \cite{lee2023electron, zhou2021perturbo, protik2022elphbolt, cepellotti2022phoebe} do not provide the functionality to include the interparticle Coulomb interactions in the collision integral. We are aware of one \textit{ab initio} study on doped silicon for which, however, the software is not publicly available \cite{caruso2016theory}. In that work, the authors include both the electron-electron and electron-plasmon interactions, albeit through a relaxation time approximation (RTA) solution of the BTE. Capturing the hydrodynamic transport physics, however, necessarily requires the full solution of the BTE, or, at least, going beyond the RTA, in order to fully capture the effect of the momentum feedback loop between the two interacting subsystems. To fill the existing methodological and knowledge gaps, we implement the necessary computational tools in the \texttt{elphbolt} \cite{protik2022elphbolt} code to enable an \textit{ab initio} study of the role of the Coulomb interactions on the charge conduction properties of real materials. Using this newly developed tool, we find that the electron-hole plasma in doped graphene can sustain three types of hydrodynamic states, one of which is the electron-hole bifluid. We find that these hydrodynamic phenomena are driven by the Coulomb interactions between the charge carriers, and that the standard phonon-limited carrier transport theory fails to capture the essential physics, both quantitatively and qualitatively. We further observe a strong Coulomb drag induced negative conductivity, corroborating recent measurements reported in Ref. \cite{ponomarenko2024extreme}. Lastly, we observe a strong violation of the Wiedemann-Franz (WF) law close to the bifluid state.

\paragraph*{Results and discussion} We operate within the linear response regime where the system is driven slightly out of equilibrium by an external electric field $\textbf{E}$ or a temperature gradient $\mathbf{\nabla} T$. The state of a particle is tagged by a band index $n_{1}$ and wavevector $\mathbf{k}_{1}$, which we write together as $1$. The non-equilibrium distribution of a state is denoted by $f_{1}$, which deviates from the equilibrium, Fermi-Dirac counterpart $f^{0}_{1}$, by a small amount. This small deviation from equilibrium is written in the following form:
\begin{equation}
    \delta f_{1} = -f^{0}_{1}(1 - f^{0}_{1})\beta(\mathbf{\nabla}T\cdot \mathbf{I}_{1} + \mathbf{E}\cdot \mathbf{J}_{1})
\end{equation}
where $\mathbf{I}_{1}$ and $\mathbf{J}_{1}$ are the response functions to the corresponding applied field and $\beta$ is the inverse temperature energy. BTEs can be written in terms of these response functions, and the solutions then give access to the various transport coefficients, expressions for which have previously been given in Ref. \cite{protik2022elphbolt}.

In this work, we focus on capturing the drag effect between electrons and holes, which is the bidirectional flow of momentum between these two systems of charge carriers under non-equilibrium conditions. Since this effect is mediated by the Coulomb interaction, it also goes by the name Coulomb drag in the literature \cite{narozhny2016coulomb}. (Generally, the Coulomb drag is defined as the drag between the charge carrier systems living on macroscopically separated materials. Graphene can be thought of as a special case where the two systems of charge carriers -- electron and holes -- coexist in the same material sample.) Now, in order to compute this effect within an \textit{ab initio} framework,  we extend the \texttt{elphbolt} code to include a screened Coulomb collision integral. This collision integral has been derived from first principles by Sanborn in Ref. \cite{sanborn1995nonequilibrium} for the case of doped, 3D semiconductors, and here we simply adapt it for our 2D case. In this approximation, the exchange interaction term is ignored. We also ignore the local field effects. In our notation, the linearized collision integral is given by
\begin{align} \label{eq:coulcoll}
    I_{1}^{\text{Coul}}[\mathbf{J}] = &\dfrac{4\pi}{\hbar f^0_{1}(1 - f^0_{1})A^{2}} \sum_{23} \sum_{\mathbf{G}} |U_{3}U^{\dagger}_{1}|^2 \notag \\ &\left|\dfrac{V^{\text{2D}, \infty}_{\mathbf{q}}(\mathbf{G})}{\epsilon^{\text{TF}}_\mathbf{q}(\mathbf{G})}\right|^{2} \delta(E_{1} + E_{2} - E_{3} - E_{4}) \notag \\
    &f^{0}_{1}f^{0}_{2}(1 - f^{0}_{3})(1 - f^{0}_{4}) \left(\textbf{J}_1 + \textbf{J}_2 - \textbf{J}_3 - \textbf{J}_4\right) \cdot \mathbf{E}, 
\end{align}
where $\hbar$ is the reduced Planck's constant, $E$ is the energy, $\mathbf{q} \equiv \mathbf{k}_{1} - \mathbf{k}_{3}$, $\hbar\omega \equiv E_{1} - E_{3}$, $A$ is the area of the crystal, $V^{\text{2D},\infty}_{\mathbf{q}}$ is the 2D bare Coulomb interaction\cite{saraga2004coulomb} screened by a high frequency dielectric, and $U^{\dagger}_{1}$ is the diagonalizer of the Wannier interpolated Hamiltonian in state $1$. Note that the wave vector of state 4 is determined by those of the remaining three. We choose the value of $\epsilon^{\infty}$ to be the out-of-plane dielectric of hBN ($4.1$)\cite{lin2013surface} to model the experimental conditions of the work in Ref. \cite{ponomarenko2024extreme}. $\epsilon^{\text{TF}}_\mathbf{q}$ is the Thomas-Fermi dielectric, given by 
\begin{equation}
     \epsilon_\mathbf{q}^{\text{TF}}( \mathbf{G}) = 1 + \dfrac{e^2 \beta}{2\varepsilon_0 A}\sum_{1}\dfrac{f^{0}_{1} (1 - f^{0}_{1})}{|\mathbf{q} + \mathbf{G}|},
\end{equation}
where $e$ is the magnitude of the electronic charge and $\varepsilon_{0}$ is the permittivity of free space. The corresponding expression for the temperature gradient field case is found through the replacement ($\mathbf{J} \rightarrow \mathbf{I}$, $\mathbf{E} \rightarrow \mathbf{\nabla} T$).

The Coulomb collision integral is added to the electron-phonon counterpart \cite{protik2022elphbolt} assuming Matthiesen's rule, i.e., that the scattering channels are independent. The final charge carrier BTE is fully solved to self-consistency using an iterative approach. That is, our approach is beyond any type of RTA. Doing so is crucial in order to accurately capture any hydrodynamics. Details of our \textit{ab initio} BTE solution method have been given before in Ref. \cite{protik2022elphbolt} and are not reproduced here. But, briefly, this approach relies on the density functional + perturbation theory as implemented in the \texttt{Quantum Espresso} code \cite{giannozzi2009quantum, giannozzi2017advanced}, maximally localized Wannier functions \cite{pizzi2020wannier90}, and the real-space, Wannier representations of the electron-phonon interaction matrix elements obtained by interfacing with the \texttt{EPW} code \cite{giustino2007electron, ponce2016epw}. Details of the employed computational methods are given in the Supplementary Information (SI) \cite{suppl}.

Since we compare to the measurements in Ref. \cite{ponomarenko2024extreme}, we include a boundary scattering term in all our calculations using a simple, phenomenological scattering rates expression: $|\mathbf{v}_{1}|/L$,
where $\mathbf{v}_{1}$ is the charge carrier group velocity and $L = 20$ $\mu\text{m}$ characterizes the sample size. This expression is added to the rest of the RTA scattering rates using, again, Matthiesen's rule. We ignore any in-scattering corrections from this scattering channel.

In this work, we choose the majority carrier concentrations to range between $1.6 \times 10^{11}$ and $1.7 \times 10^{12}$ cm$^{-2}$. Getting any closer to the CNP is not possible at the moment due to the extreme computational complexity; see SI \cite{suppl}. In any case, the application of the semiclassical BTE is invalid near the CNP, as mentioned in the introduction. 

\begin{figure}
    \centering
    \includegraphics[width=1.0\linewidth]{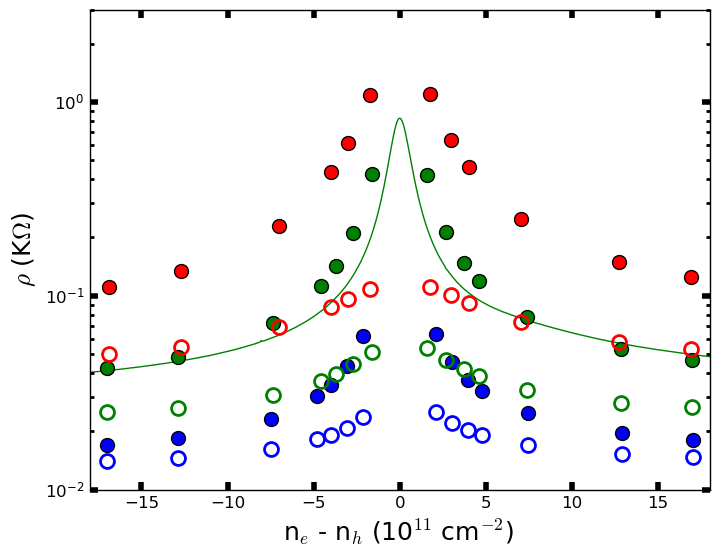}
    \caption{Resistivity vs. charge carrier concentration at various temperatures. Positive (negative) sector of the horizontal axis corresponds to majority electron (hole) concentration. Solid green line gives the 300 K measurements from Ref. \cite{ponomarenko2024extreme} (data obtained from the primary author via private communication and converted to resistivity units). Calculated values are presented in disks for three temperatures: 200 K (blue), 300 K (green), and 400 K (red). Solid (empty) symbols correspond to the theory that includes (excludes) the Coulomb interactions.}
    \label{fig:resistivity} 
\end{figure}

In Fig. \ref{fig:resistivity}, we present our calculated room temperature resistivity as a function of the difference of the electron and hole concentrations, $n_{e} - n_{h}$. We include a comparison to recent $300$ K measurements from Ref. \cite{ponomarenko2024extreme} (green line). We find that without the inclusion of the Coulomb interactions (empty disks), the calculated resistivity is in poor agreement with the measurements (green line vs. green empty disks) over the entire range of concentrations considered. The theory including the Coulomb interactions gives quantitatively good agreement (green line vs. green solid disks) in the moderate to high concentration range. For low concentrations, however, the Coulomb enabled theory overestimates the resistivity. This could be due to a couple of reasons. First, the lattice constant used in our theoretical modeling might not match the one of the sample in Ref. \cite{ponomarenko2024extreme} since the graphene flake there is sandwiched between hBN substrates. We take into account the role of hBN only through an extra screening factor, as mentioned earlier. Secondly, our Coulomb collision integral includes only the direct interaction term. Our theoretical predictions might be improved through the inclusion of the exchange term, since the direct and exchange terms have opposite signs. We leave this for a future work. Now, the transport coefficients presented in Fig. \ref{fig:resistivity} being band summed quantities, hide certain crucial failures of the Coulomb-free theory. And in order to observe these failures of the phonon limited transport to capture important microscopic physics, we next look at the valence and conduction band contributions separately.

\begin{figure}
    \centering
    \includegraphics[width=1.0\linewidth]{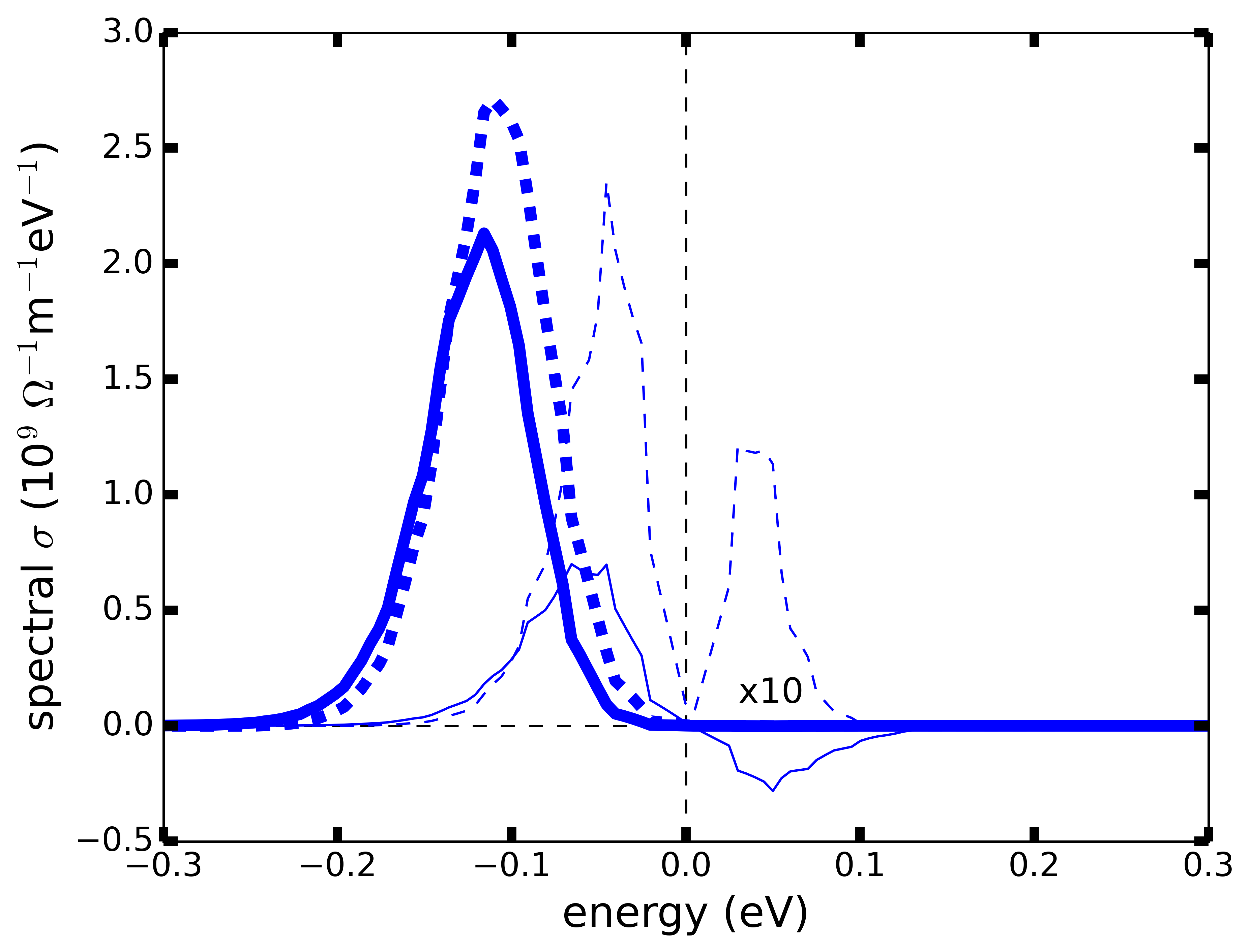}
    \caption{$200$ K spectral conductivity from the valence and conduction bands for hole doped cases. Thick (thin) lines are for a majority hole concentration of $1.3\times10^{12}$ ($2\times10^{11}$) cm$^{-2}$. Solid (dashed) lines denote the theory including (excluding) the Coulomb interactions. The Dirac point is set at zero energy. The contribution from the conduction band (positive energy) is displayed with a factor $10$ enlargement.}
    \label{fig:speccond}
\end{figure}

In Fig. \ref{fig:speccond}, we give the spectral conductivities at $200$ K for a high (thick lines) and a low (thin lines) hole doped case. The Dirac point is placed at the zero of the horizontal axis. Note that the conduction band contribution is plotted with a scaling factor of $10\times$. For the Coulomb-free case (dashed lines), both the valence and the conduction bands give positive contributions to the conductivity, as one would expect from a classical Drude picture. In stark contrast, the activation of the Coulomb interactions (solid lines) leads to a significantly lower positive contribution from the valence band, along with a \textit{negative} contribution from the conduction band. A completely analogous effect happens for the electron-doped case, where the inclusion of the Coulomb interactions causes the holes to give a negative contribution to the conductivity; see SI \cite{suppl}. There we also show that this effect persists even at room temperature. What we described so far is the evidence for an extreme Coulomb drag effect between the electron and hole systems. The essential physics we observe here is that the majority carriers transfer momentum to the minority ones, causing them to develop a current in a direction opposite to what is dictated by the coupling to the external electric field. On the other hand, the minority carriers also drag the majority carriers in the opposite direction, causing them to give a lower, albeit still positive, contribution. This extreme Coulomb drag effect becomes more prominent the closer one gets to the CNP. Note that this phenomenon has been experimentally observed in Ref. \cite{ponomarenko2024extreme}. It is now interesting to inquire whether the mode of this charge transport is hydrodynamic. This question is motivated by the prevailing wisdom that the Coulomb scatterings between the charge carriers contain a large proportion of the momentum conserving (Normal) type processes. While it is well-known that in 2D systems the decay of the screened Coulomb potential goes as $r^{-1}$ for small distance $r$ as opposed to exponentially (Yukawa-type) as in 3D, making the former a long-range interaction \cite{Giuliani_Vignale_2005}. However, having a long-range Coulomb interaction alone is not a sufficient condition for the hydrodynamic behavior. The scattering events must also be predominantly of the Normal type.
In such a case, the non-equilibrium distribution function takes the form of a shifted Fermi distribution, the shift amount being characterized by a global drift velocity. We investigate this next.

\begin{figure*}
    \centering
    \begin{subfigure}{0.5\textwidth} 
        \centering
        \includegraphics[width=0.9\textwidth]{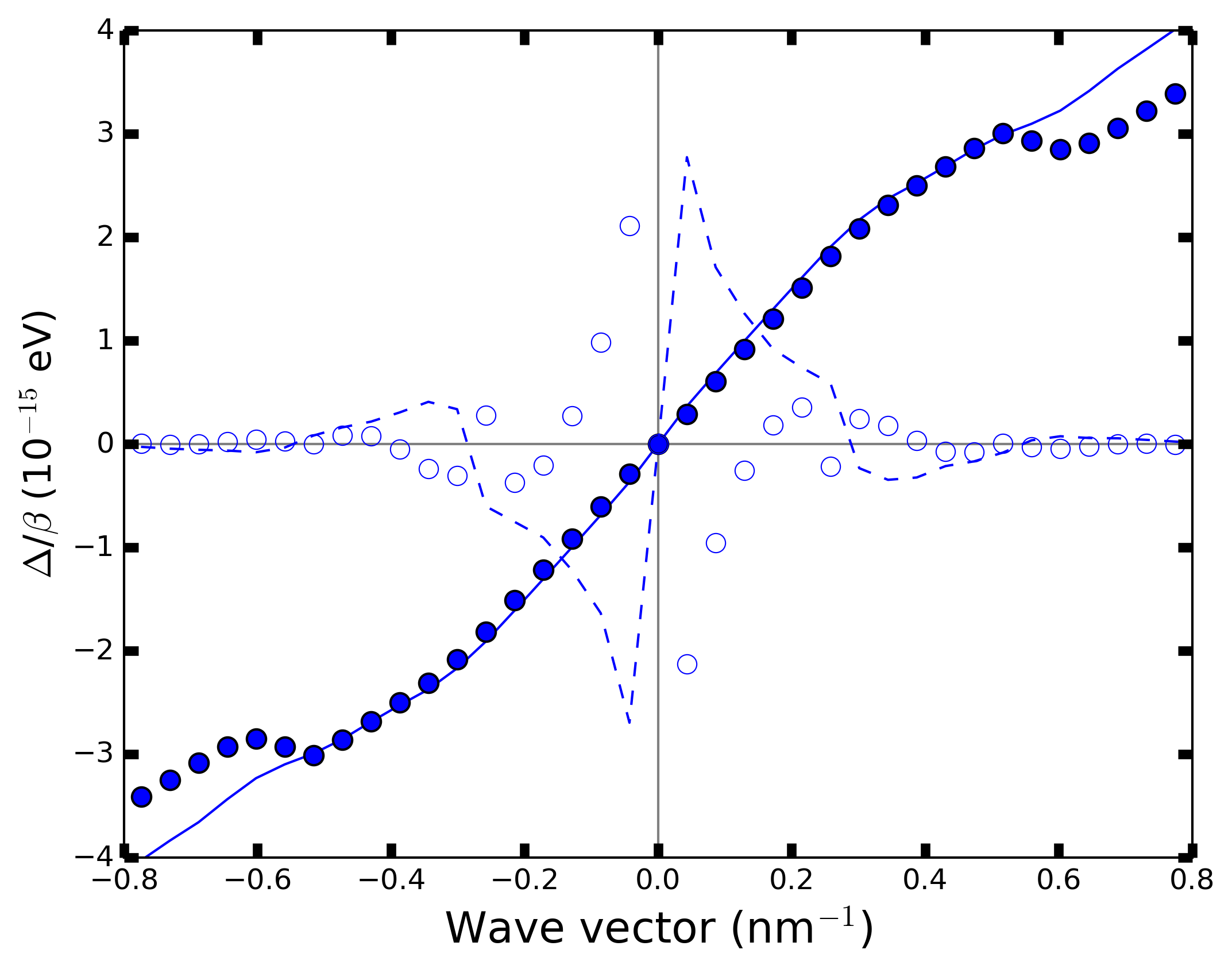}
        \caption{T = $200$ K, $n_{h} - n_{e}$ = $2.1\times10^{11}\text{ cm}^{-2}$}
    \end{subfigure}\hfill 
    \begin{subfigure}{0.5\textwidth}
        \centering
        \includegraphics[width=0.9\textwidth]{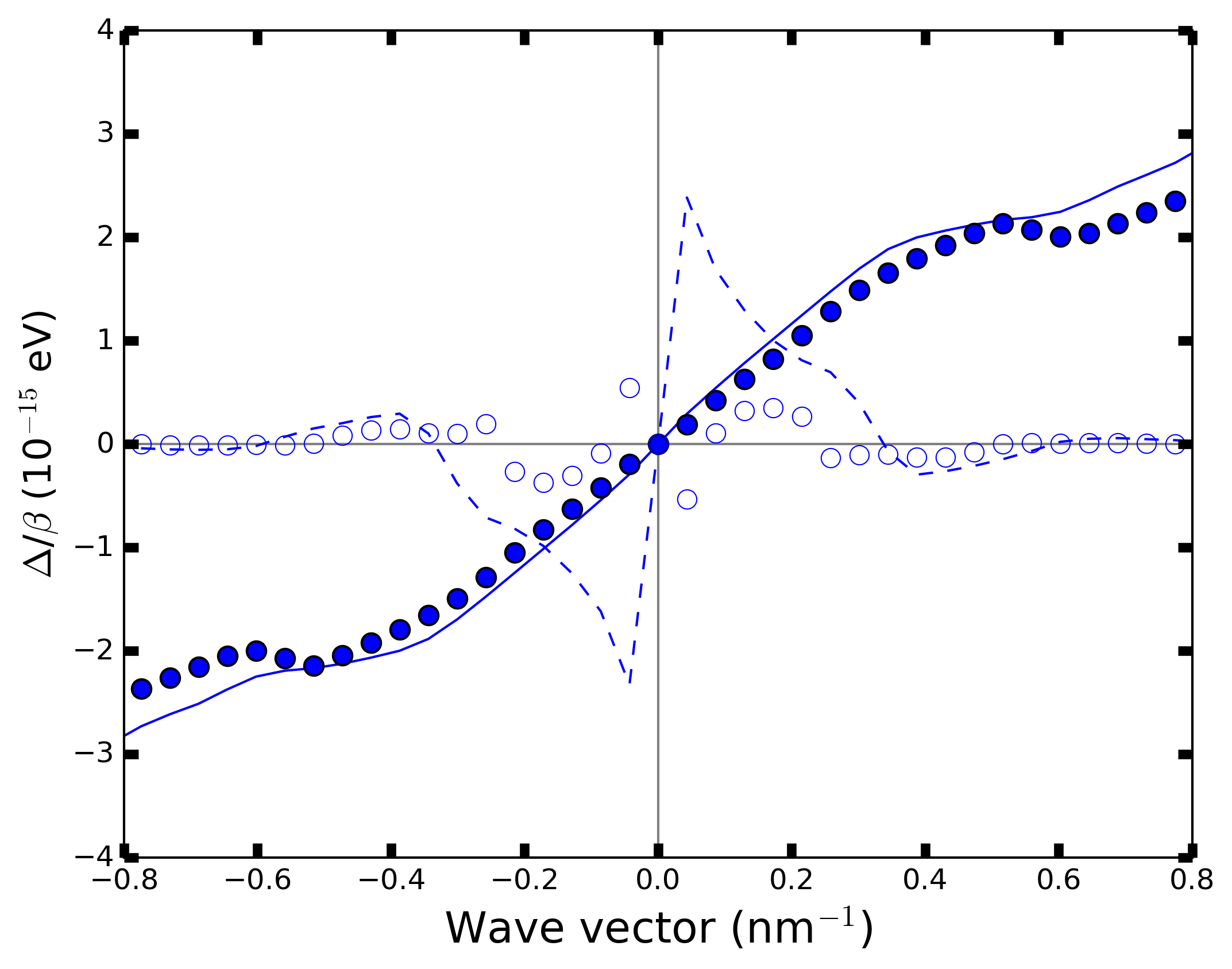}
        \caption{T = $200$ K, $n_{h} - n_{e}$ = $7.4\times10^{11}\text{ cm}^{-2}$}
    \end{subfigure}
    \vspace{1em} 
    \begin{subfigure}{0.5\textwidth}
        \centering
        \includegraphics[width=0.9\textwidth]{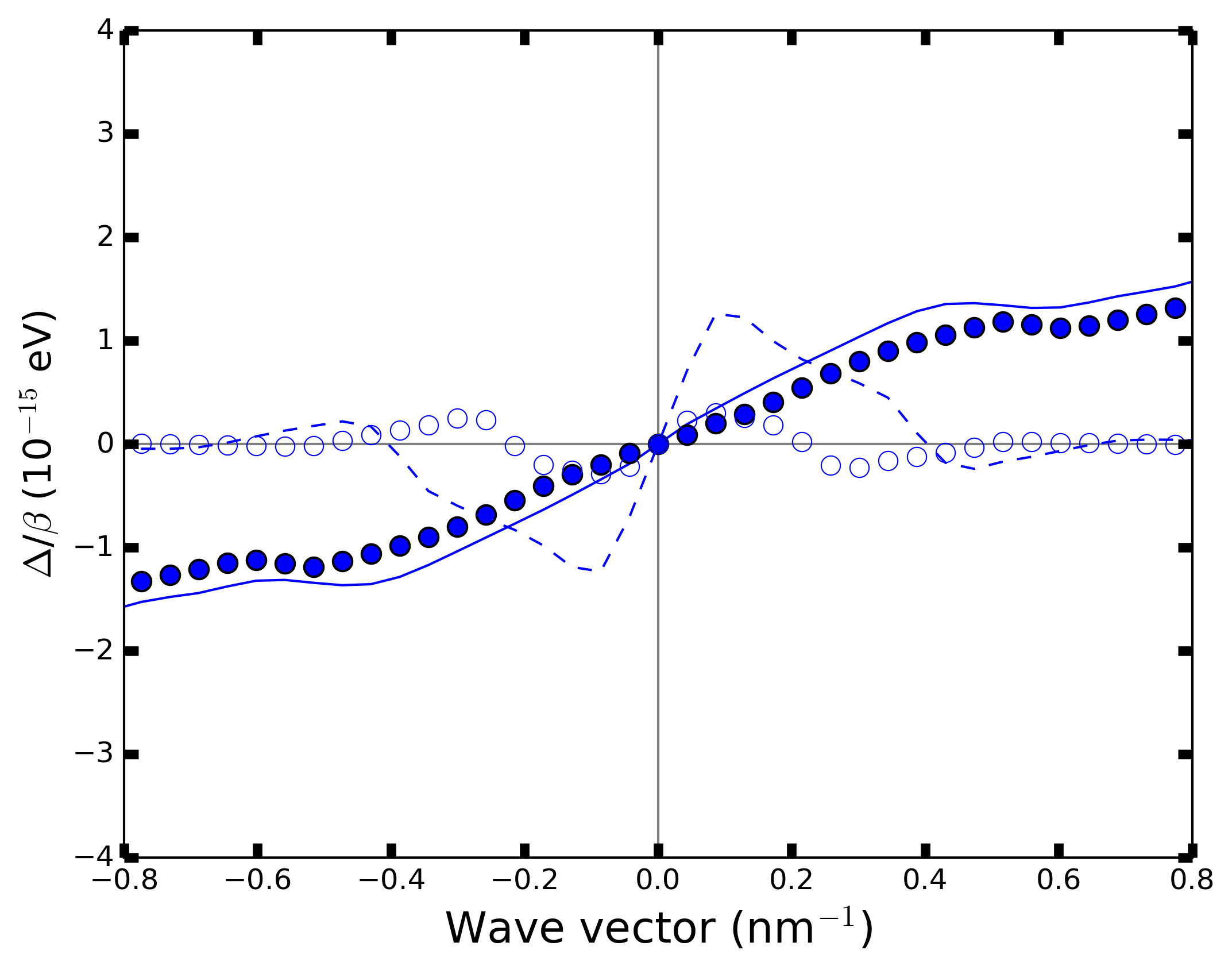}
        \caption{T = $200$ K, $n_{h} - n_{e}$ = $1.7\times10^{12}\text{ cm}^{-2}$}
    \end{subfigure}\hfill
    \begin{subfigure}{0.5\textwidth}
        \centering
        \includegraphics[width=0.9\textwidth]{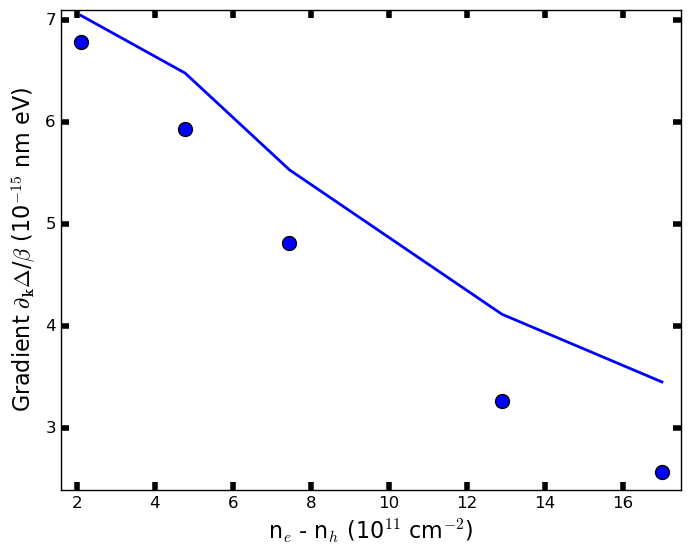}
        \caption{T = $200$ K}
    \end{subfigure}
    \caption{\parbox{\textwidth}{Panels \textbf{a}, \textbf{b}, \textbf{c}: Normalized deviation function $\Delta/\beta$ vs. magnitude of electron wavevector, measured from \textbf{K} point and along the (\textbf{K}-$k_x$)--\textbf{K}--(\textbf{K}+$k_x$) direction of the Brillouin zone. Panel \textbf{d}: Gradient $\partial_{\mathbf{k}}\Delta$ near the $\mathbf{K}$-point for the valence and conduction band. In all plot, lines represent the valence band and disks represent the conduction band. Solid lines and symbols are for the Coulomb enabled theory, whereas dashed lines and empty symbols are for the Coulomb-free theory.}}
    \label{fig:deviation}
\end{figure*}

In Fig. \ref{fig:deviation}, panels a, b, and c, we plot for the $200$ K case the band resolved, normalized deviation function $\Delta_{1}/\beta$ where $\Delta_{1} \equiv \delta f_{1}/f^{0}_{1}/(1 - f^{0}_{1})$. (This definition is the Fermionic analogue of the Bosonic one given in Ref. \cite{lee2015hydrodynamic}.) We do this along a section of the  (\textbf{K}-$k_x$)--\textbf{K}--(\textbf{K}+$k_x$) path of the Brillouin zone measured from the \textbf{K} point, for three different majority hole doped cases. Here $k_{x}$ is proportional to the unit vector in the $x$ direction. The electric field is set at $1$ eV/nm/C, in both $x$ and $y$ direction. We observe that this function is strongly linear, either for one or both of the electron and hole subsystems, only when the Coulomb interactions are turned on (solid disks and solid lines). Note that the linearity of $\Delta_{1}$ is the signature of the hydrodynamic transport regime. We notice that, as we go from the high doped (panel c, furthest from charge neutrality) toward low doped (panel a, closest to charge neutrality), both the valence (solid line) and conduction (solid disks) bands tend toward having a robustly linear $\Delta_{1}$. That is, by tuning the doping toward charge neutrality, both the electron and hole subsystems go hydrodynamic. On the other hand, as we move further away from the charge neutrality condition, we retain robust hydrodynamics in the holes but begin to lose it in the electrons. Now, the most interesting result is that in the low doped limit, panel a, the electron and hole subsystems tend toward a \textit{shared} hydrodynamic state -- the electron-hole bifluid. To quantify the last statement, we plot in panel d the gradient $\partial_{\mathbf{k}}\Delta_{1}$ taken near the \textbf{K} point, along the same Brillouin zone path. We see that the slopes for both the valence and the conduction bands approach a common value in the low doping limit. This is most pronounced for the low temperature case. A completely analogous situation occurs for the majority electron doped cases, as shown in the SI \cite{suppl}. We also show there that, with increasing temperature, this phenomenon becomes less pronounced. The bottom line is that the electron-hole bifluidity in doped graphene manifests in the low temperature and low doping situations. The simpler theory that lacks the Coulomb interactions does not capture this essential physics (dashed lines and empty disks).

\begin{figure}
    \centering
    \includegraphics[width=1.0\linewidth]{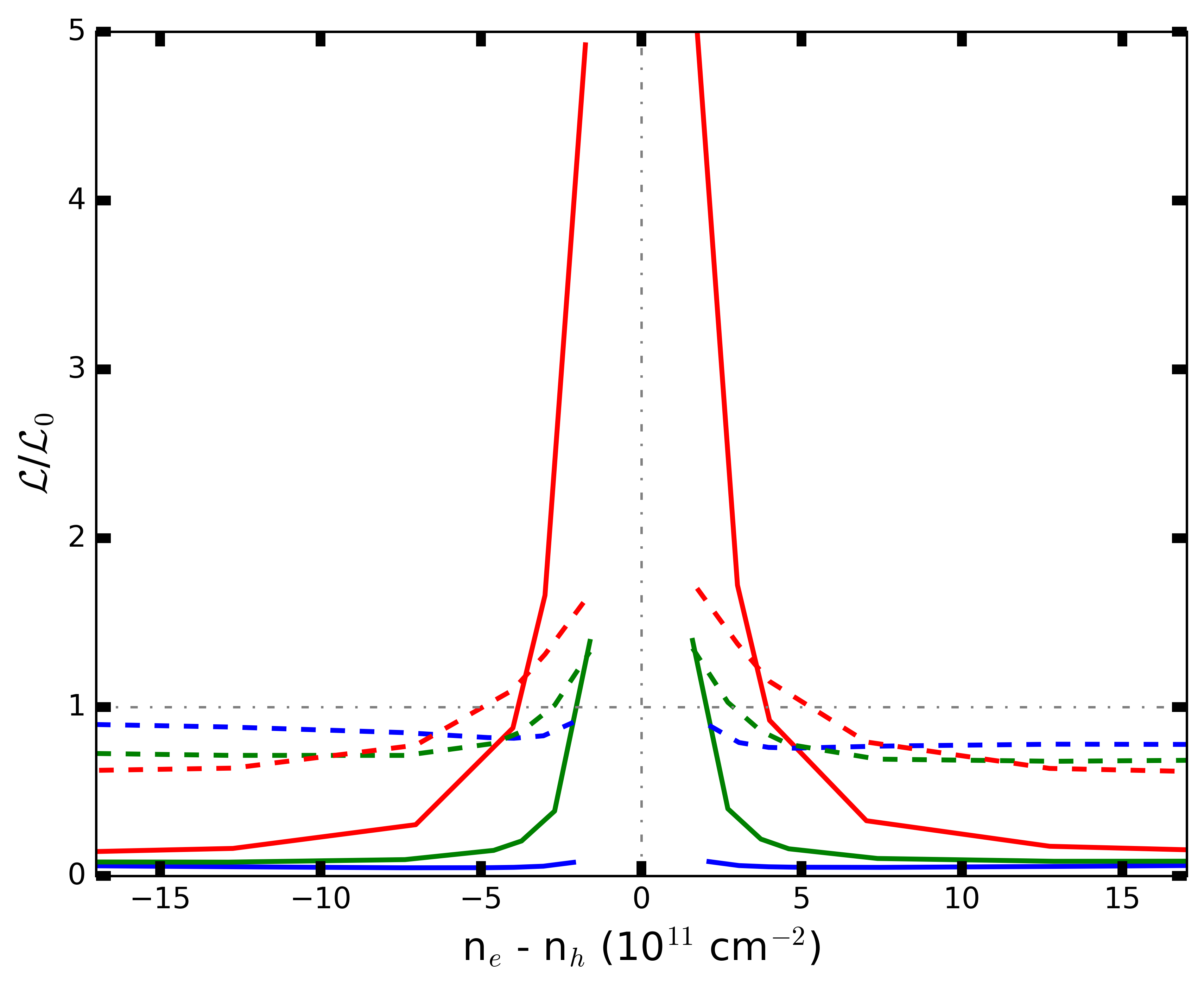}
    \caption{Scaled Lorentz number vs. carrier concentration. Blue, green, and red symbols are for $200$, $300$, and $400$ K, respectively. Solid (broken) lines correspond to the theory including (excluding) the Coulomb interaction. The scaling is with respect to the universal value ($\mathcal{L}_{0} \equiv 2.44 \times 10^{-8} \text{ V}^2 \text{ K}^{-2}$) according to the WF law.}
    \label{fig:lorentz}
\end{figure}

Finally, in Fig. \ref{fig:lorentz}, we plot the Lorentz number $\mathcal{L} = \kappa_{\text{el}}(\sigma T)^{-1}$ scaled by the universal (Sommerfeld) value $\mathcal{L}_{0}$ as a function of $n_{e} - n_{h}$, for three temperature points. Here $\kappa_{\text{el}}$ is the electronic thermal conductivity in the open circuit condition and $\sigma$ is the charge conductivity. The prevailing wisdom is that very close to the CNP in graphene at low temperatures, a violation of the WF law occurs due to the domination of the electron-hole interactions, which results in the thermal resistivity to vanish for a finite electric current \cite{crossno2016observation}. Here we find that when the Coulomb interactions are included (solid blue, green, and red lines), there is a strong violation of the WF law even in the electron and hole doped regimes. However, this violation increases as we approach the charge neutrality, consistent with the conclusion made in Ref. \cite{crossno2016observation}. This, however, does not correlate with the bifluidity, which we found to be a dominant effect at low temperatures, while here, the WF law violation increases at high temperatures. The theory that ignores the Coulomb interactions (dashed  blue, green, and red lines), also shows a violation of the WF law, though much weaker than what is predicted by the better theory.

\paragraph*{Summary and outlook} In this work we implement an \textit{ab initio} method to include the Coulomb interactions among the charge carriers within the \texttt{elphbolt} Boltzmann transport codes suite. We use this new tool to compute the charge carrier transport properties of doped graphene. We find that the Coulomb interaction enabled theory predicts three hydrodynamic transport regimes: (1) exclusive hole hydrodynamics in the high hole doped case, (2) exclusive electron hydrodynamics in the high electron doped case, and (3) coupled electron-hole bilfluidity in the weak hole or electron doped cases at low temperature. Our results corroborate the recent observations of negative conductivity in the extreme Coulomb drag regime \cite{ponomarenko2024extreme}. We also find a strong violation of the WF law that is driven by the Coulomb interactions between electrons and holes. None of the above mentioned phenomena is captured by the theory that includes the electron-phonon interactions, but ignores Coulomb.

Although the \textit{ab initio} methodology described in this work is able to predict the strong Coulomb drag driven negative conductivity and the electron-hole bifluidity, we see several avenues for improvement and extensions that we wish to touch upon in future works. We briefly discuss them below.
\begin{itemize}
    \item Exchange term: As mentioned earlier, our collision integral contains only the direct term. Since the direct and exchange terms come with opposite signs, we think that the inclusion of the latter will reduce calculated resistivity values at low doping levels, and result in theoretical predictions that are in better agreement with measurements.

    \item Dynamical screening: In this work, we employ the Thomas-Fermi screening model. One can, in fact, use a dynamical screening model within the random phase approximation (RPA) \cite{sanborn1995nonequilibrium}. The RPA includes the effect of the plasmons in the Coulomb kernel and, in principle, captures the screening better than the static Thomas-Fermi theory.

    \item Electron-phonon drag: In this work, we assume that the phonon system can be taken to remain in equilibrium. We do this for two reasons. Firstly, we notice that the transport physics in the doping and temperature ranges considered here is entirely dictated by the Coulomb interactions, as is evidenced from the failure of the theory that ignores these. Secondly, computing the phonon thermal conductivity in graphene is a difficult and contentious matter. It has been argued that including the extremely computationally demanding four-phonon interactions is necessary to make decent lattice thermal conductivity predictions. That said, there might as well be other systems where the extremely expensive four-phonon interactions do not have to be considered. In such system, it might be possible to carry out a coupled electron-phonon BTEs calculation including the Coulomb interactions. We note that in Ref. \cite{quan2025coupled}, the authors predict a joint electron-phonon hydrodynamics in doped MoS$_{2}$. An interesting research question would then be whether there exist materials that, under suitable conditions, exhibit both the electron-hole and electron-phonon hydrodynamics.
\end{itemize}

\paragraph*{Acknowledgments} This work was funded by the Deutsche Forschungsgemeinschaft (DFG, German Research
Foundation) through the Emmy Noether research grant
(Grant No. 534386252). This work used computational resources provided by Noctua2 at the Paderborn Center for Parallel Computing (PC2), JUWELS at the Jülich Supercomputing Centre, and the CMS cluster of Humboldt-Universität zu Berlin. We thank Dr. Leonid A. Ponomarenko for kindly providing the resistivity measurement from Ref. \cite{ponomarenko2024extreme}.

\paragraph*{Data and code availability} The theory results supporting the findings of this work will be made available at the time of publication. The 
\texttt{elphbolt} code used to carry out the calculations is free/libre and open source software.

\bibliography{biblio}
\end{document}


\title{Supplementary Information (SI) for Coulomb drag driven electron-hole bifluidity in doped graphene}

\author{Dwaipayan Paul}
\email{Contact author: dwaipayan.paul@hu-berlin.de}

\author{Elena Trukhan}
\email{Contact author: elena.trukhan@physik.hu-berlin.de}

\author{Nakib H. Protik}%
\email{Contact author: nakib.protik@physik.hu-berlin.de}
\affiliation{%
 Institut f\"{u}r Physik and CSMB, Humboldt-Universit\"{a}t zu Berlin
}

\date{\today}

\maketitle

\section{Computational details}
\subsection{Electrons, phonons, and Wannierization}
We use the \texttt{Quantum Espresso} \cite{giannozzi2009quantum}\cite{giannozzi2017advanced}\cite{lee2023electron} code suite for our density functional theory (DFT) and density functional pertubation theory (DFPT) calculations. We choose a norm-conserving pseudopontential, with the Perdew-Zunger (local density approximation) exchange-correlation \cite{perdew1981self}, and use a plane-wave energy cutoff of 60 eV. Our calculated relaxed lattice constant is $2.434$ \AA, which is close to the literature value of 2.46 \AA \cite{razado2018structural}. We choose a large vacuum size of $\sim 8.5 $\AA, to eliminate any effect of periodic boundary condition in the DFT/DFPT calculations.

To compute phonons, we use a $36 \times 36 \times 1$/$18 \times 18 \times 1$ electron/phonon wave vector ($\mathbf{k}/\mathbf{q}$) set of meshes.

To generate Wannierized quantities, we use the \texttt{EPW} \cite{giustino2007electron}\cite{ponce2016epw} module of the \texttt{Quantum Espresso} suite. We use a coarse $18 \times 18 \times 1$ for both $\mathbf{k}$ and $\mathbf{q}$ meshes. We Wannierize 5 bands, choosing $sp^2$ and $p_z$ as the initial projection on one of the basis atoms, and $p_z$ on the other.
\begin{figure*}[ht!]
    \centering
    \begin{subfigure}{0.5\textwidth} 
        \centering
        \includegraphics[width=0.9\textwidth]{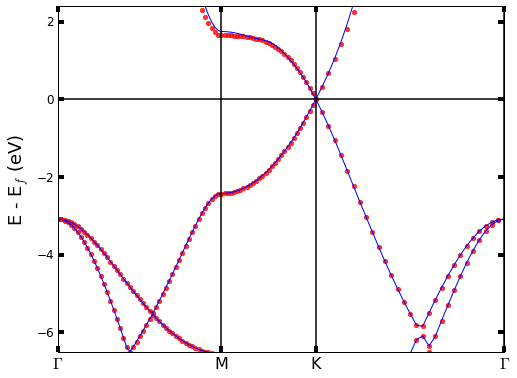}
        \caption{}
    \end{subfigure}\hfill 
    \begin{subfigure}{0.5\textwidth}
        \centering
        \includegraphics[width=0.9\textwidth]{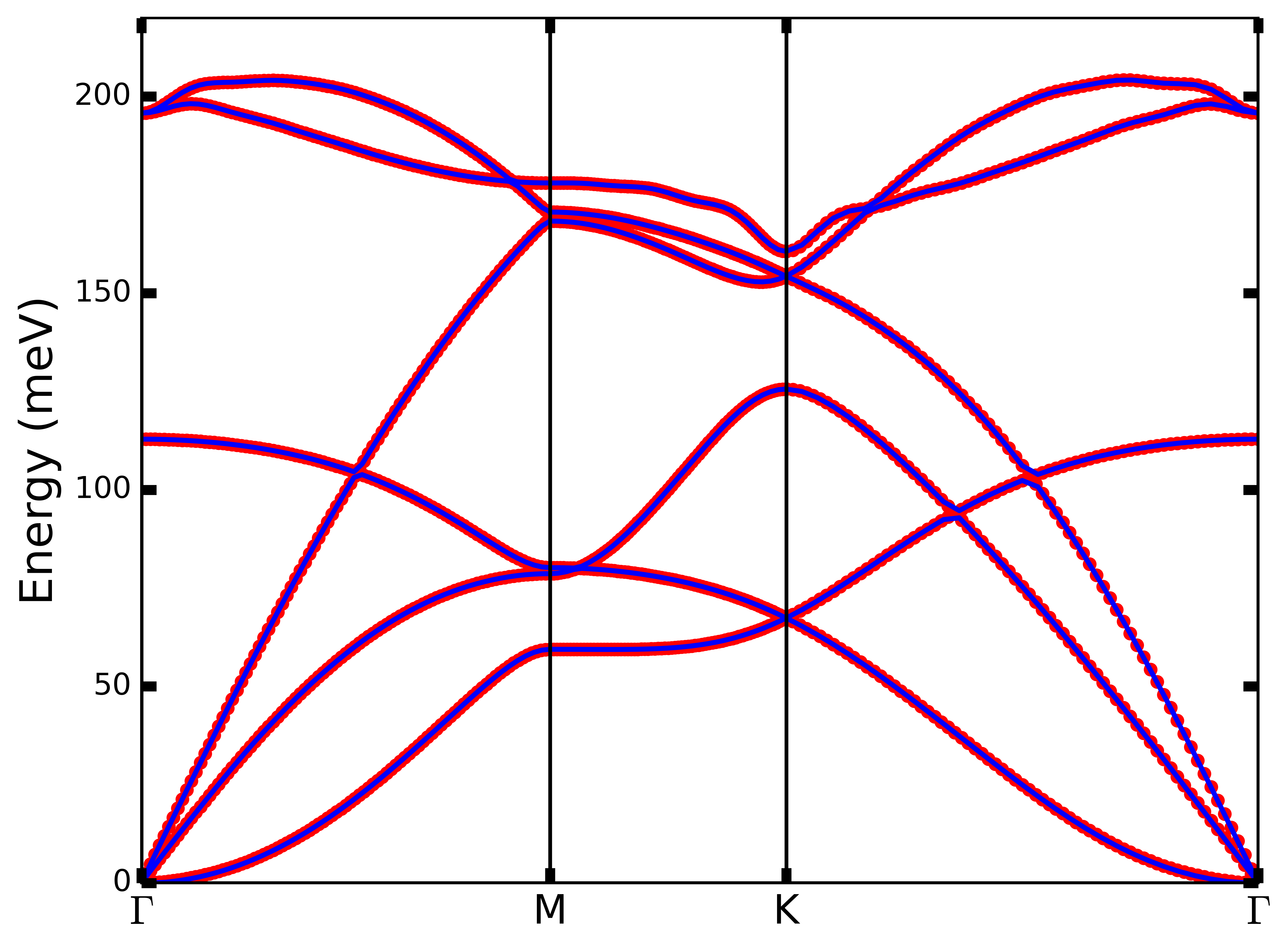}
        \caption{}
    \end{subfigure}
    \caption{\parbox{\textwidth}{Comparison of electron (panel \textbf{a}) and phonon (panel \textbf{b}) dispersions from DFT and DFPT, compared to the corresponding Wannier interpolation. The electron energies are measured with respect to the Fermi level $E_{f}$.}}
    \label{fig:epw}
\end{figure*}

\begin{figure*}[ht!]
    \centering
    \begin{subfigure}{0.5\textwidth} 
        \centering
        \includegraphics[width=0.9\textwidth]{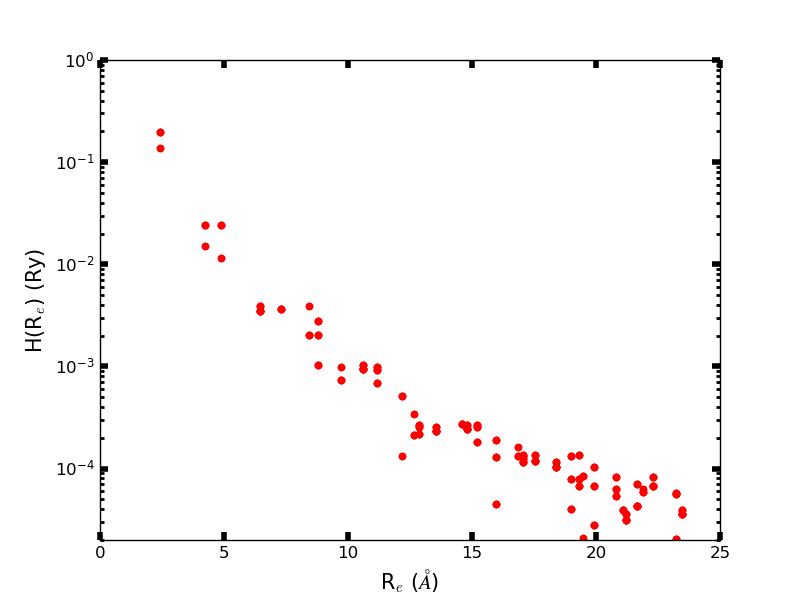}
        \caption{}
    \end{subfigure}\hfill 
    \begin{subfigure}{0.5\textwidth}
        \centering
        \includegraphics[width=0.9\textwidth]{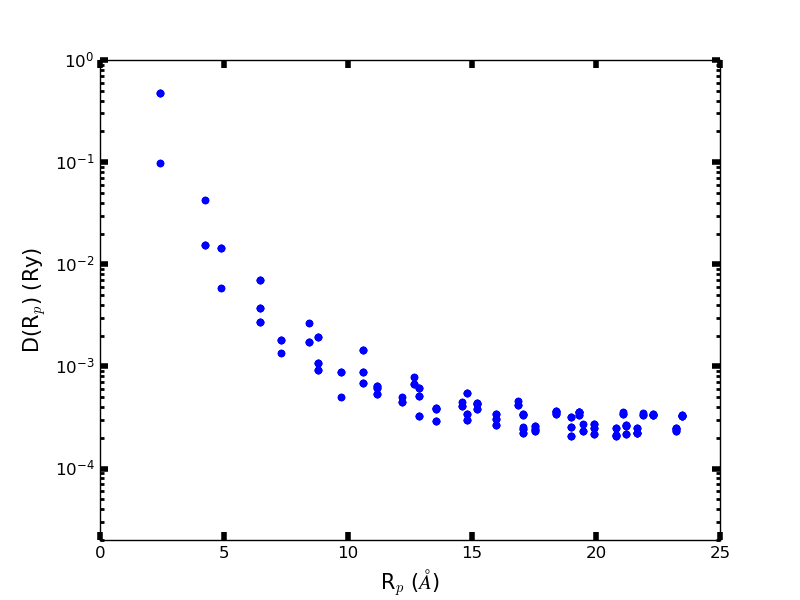}
        \caption{}
    \end{subfigure}
    \begin{subfigure}{0.5\textwidth} 
        \centering
        \includegraphics[width=0.9\textwidth]{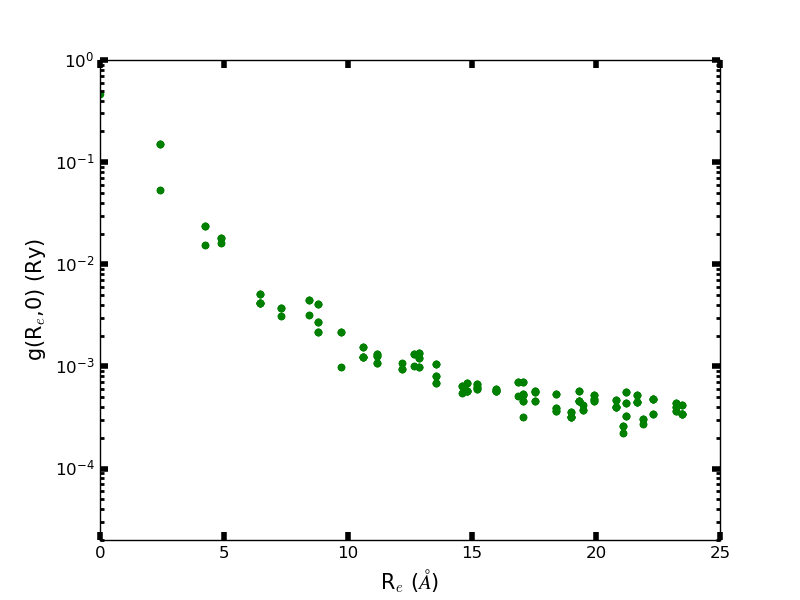}
        \caption{}
    \end{subfigure}\hfill 
    \begin{subfigure}{0.5\textwidth}
        \centering
        \includegraphics[width=0.9\textwidth]{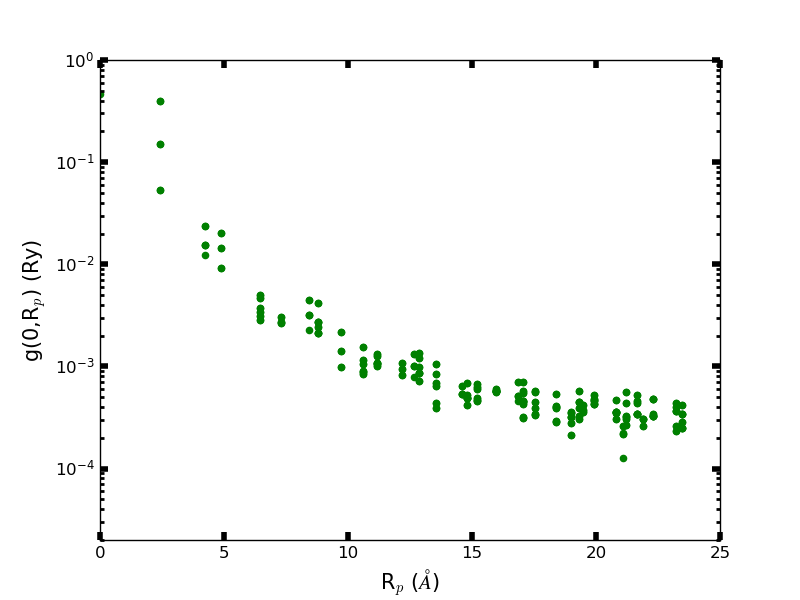}
        \caption{}
    \end{subfigure}
    \caption{\parbox{\textwidth}{Real-space decay plots of electronic Hamiltonian (panel \textbf{a}), phonon dynamical matrix (panel \textbf{b}) and electron-phonon matrix elements (panel \textbf{c} and \textbf{d})}}
    \label{fig:epwdecay}
\end{figure*}

In Fig. \ref{fig:epw}, we show a comparison of the direct DFT/DFPT electron/phonons and the corresponding Wannier interpolated ones. Note that only those states within a thin shell around the chemical potential contributes to transport, and in our calculations we only lightly dope the system. In this energy range, the Wannier interpolated energies are in very good agreement with the direct DFT ones. We show further in Fig. \ref{fig:epwdecay} the decay of the Wannierized electronic Hamiltonian, phonon dynamical matrix, and the electron-phonon matrix elements with respect to the distance in real space ($R_{e}$ for the electron grid and $R_{p}$ for the phonon grid). We use a modified version of the \texttt{EPW} given in Ref. \cite{githubGitHubNakibelphbolt} that prints out the necessary quantities that would later be needed in the transport code, \texttt{elphbolt} \cite{protik2022elphbolt}.

\subsection{Transport convergence and computational complexity}
  \begin{figure}[ht!]
     \centering
     \includegraphics[width=0.6\columnwidth]{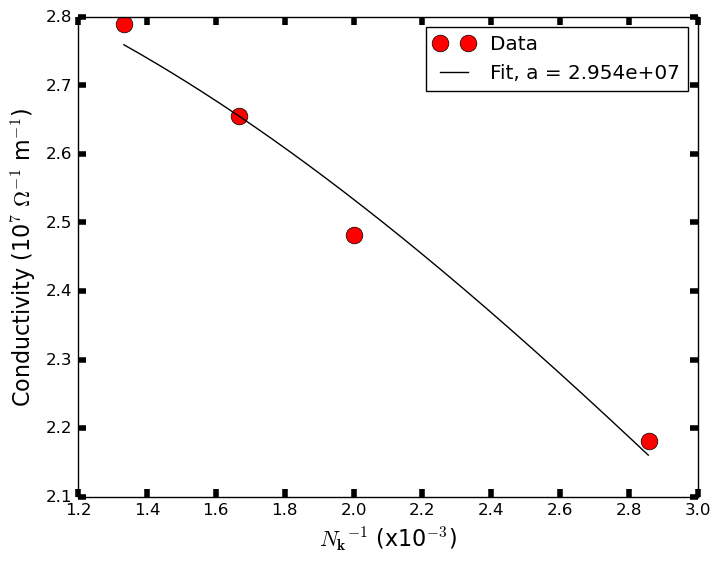}
     \caption{Conductivity vs. inverse of number of wave vectors in the transport mesh. Temperature is set to $300$ K and hole concentration is set to $4.8\times10^{11}$ cm$^{-2}$. The vertical intercept of the fitted Gaussian is denoted by $a$.}
     \label{convg}
 \end{figure}

 \begin{table}[h!]
    \centering
    \begin{tabular}{|c|c|c|c|c|}
      \hline
      $\mathbf{k}$-mesh   & $350\times350$ & $500\times500$ & $600\times600$ & $750\times750$  \\\hline
        No Coulomb & $0.068$ & $0.081$ & $0.12$ & $0.22$\\ \hline
        With Coulomb & $148.21$ & $1381.58$ & $4584.27$ & $15866.59$\\ \hline
    \end{tabular}
    \caption{Total time needed for self-consistency in the iterative electronic BTE. Time given in CPU-hours. These calculations were carried out on the Noctua2 cluster at the Paderborn Center for Parallel Computing (PC2), for a hole concentration of $4.8\times10^{11}$ cm$^{-2}$ at $300$ K.}
    \label{tab:timing}
\end{table}

Since \texttt{elphbolt} uses the triangular method \cite{kurganskii1985integration} for the evaluation of the energy conserving $\delta$-functions, we only have to converge our transport properties with respect to the $\mathbf{k}$-mesh. Note that the $\textbf{k} + \mathbf{q}$-mesh is identical to the $\mathbf{k}$-mesh in our method. For a number of ($N_{\mathbf{k}}\times N_{\mathbf{k}}\times 1$) mesh densities, we plot the conductivity as a function of the inverse of the number of points in the $\mathbf{k}$-mesh, $N_{\mathbf{k}}^{-1}$. This set of calculations is done at 300 K, with a hole carrier concentration of $4.8\times10^{11}$ cm$^{-2}$. We include both the electron-phonon and Coulomb interactions. We fit the calculated data to a Gaussian of the form $ae^{-bx^2}$. The intercept of the vertical axis gives the $\infty$-mesh density limit, which comes out to be $2.954 \times 10^7$ $\Omega^{-1}$ m$^{-1}$. The conductivity given by the $600 \times 600 \times 1$ $\mathbf{k}$-mesh is within $\approx 10 \%$ of this value, and the one given by the $750 \times 750 \times 1$ $\mathbf{k}$-mesh is within $\approx 5 \%$. All results presented in the main text are computed with the $600 \times 600 \times 1$ mesh.

We choose a transport active Fermi window of $1.2$ eV with the Dirac point placed at the center.

We give the computational timings for the full solution of the electronic BTE for various meshes in Tab. \ref{tab:timing}. We include both the expensive (Coulomb activated) and the significantly cheaper (Coulomb free) calculations.

\section{Spectral conductivity for electron doped graphene}
\begin{figure*}[ht!]
    \centering
    \begin{subfigure}{0.5\textwidth} 
        \centering
        \includegraphics[width=0.9\textwidth]{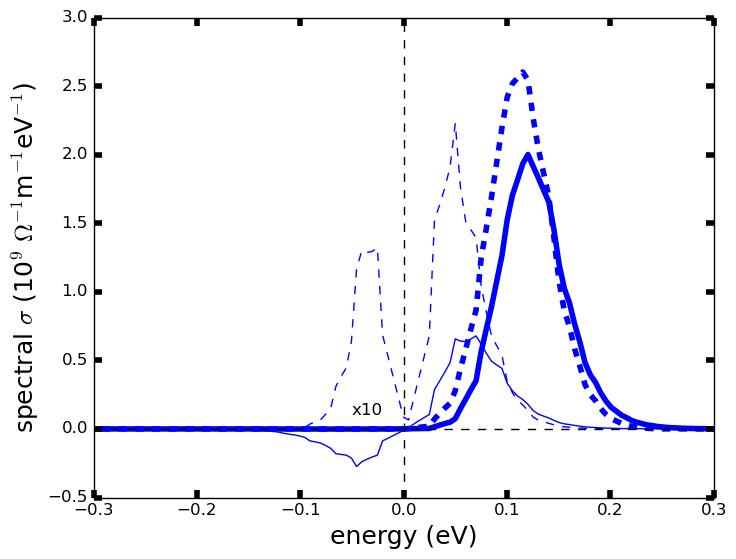}
        \caption{}
    \end{subfigure}\hfill 
    \begin{subfigure}{0.5\textwidth}
        \centering
        \includegraphics[width=0.9\textwidth]{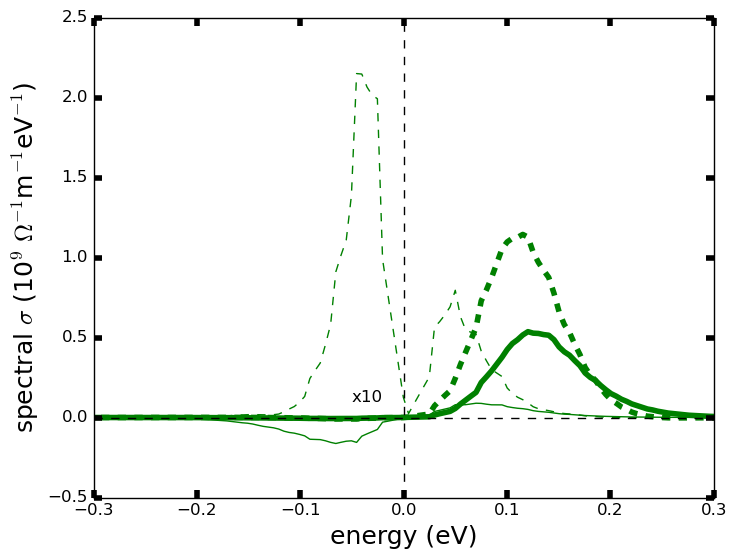}
        \caption{}
    \end{subfigure}
    \vspace{1em} 
    \begin{subfigure}{0.5\textwidth}
        \centering
        \includegraphics[width=0.9\textwidth]{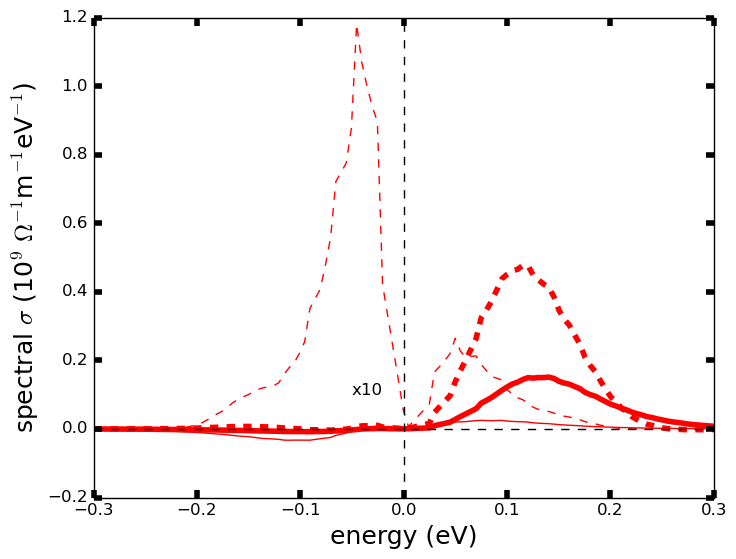}
        \caption{}
    \end{subfigure}
    \caption{\parbox{\textwidth}{Spectral conductivity from the valence and conduction bands for electron doped cases for temperatures $200$ K (\textbf{a}), $300$ K (\textbf{b}),  $400$ K (\textbf{c}). Thick (thin) lines are for a majority electon concentration of $1.3\times10^{12}$ ($2\times10^{11}$) cm$^{-2}$. Solid (dashed) lines denote the theory including (excluding) the Coulomb interactions. The Dirac point is set at zero energy. The contribution from the valence band (negative energy) is displayed with a factor $10$ enlargement.}}
    \label{fig:spec_cond}
\end{figure*}
In Fig. \ref{fig:spec_cond}, we provide the spectral conductivities for high (thick lines) and low (thin lines) electron doped graphene for 3 different temperature points - $200$ K (blue), $300$ K (green) and $400$ K (red). We saw earlier in the main text how in the hole-doped case, the conduction states can give negative contributions to the conductivity when the Coulomb interactions are included. We see the same here (solid lines), but for the valence states. For the case without the Coulomb interactions (dashed line), this is effect is absent.

\section{Normalized deviation function for electron doped graphene}
\begin{figure*}[ht!]
    \centering
    \begin{subfigure}{0.5\textwidth} 
        \centering
        \includegraphics[width=0.9\textwidth]{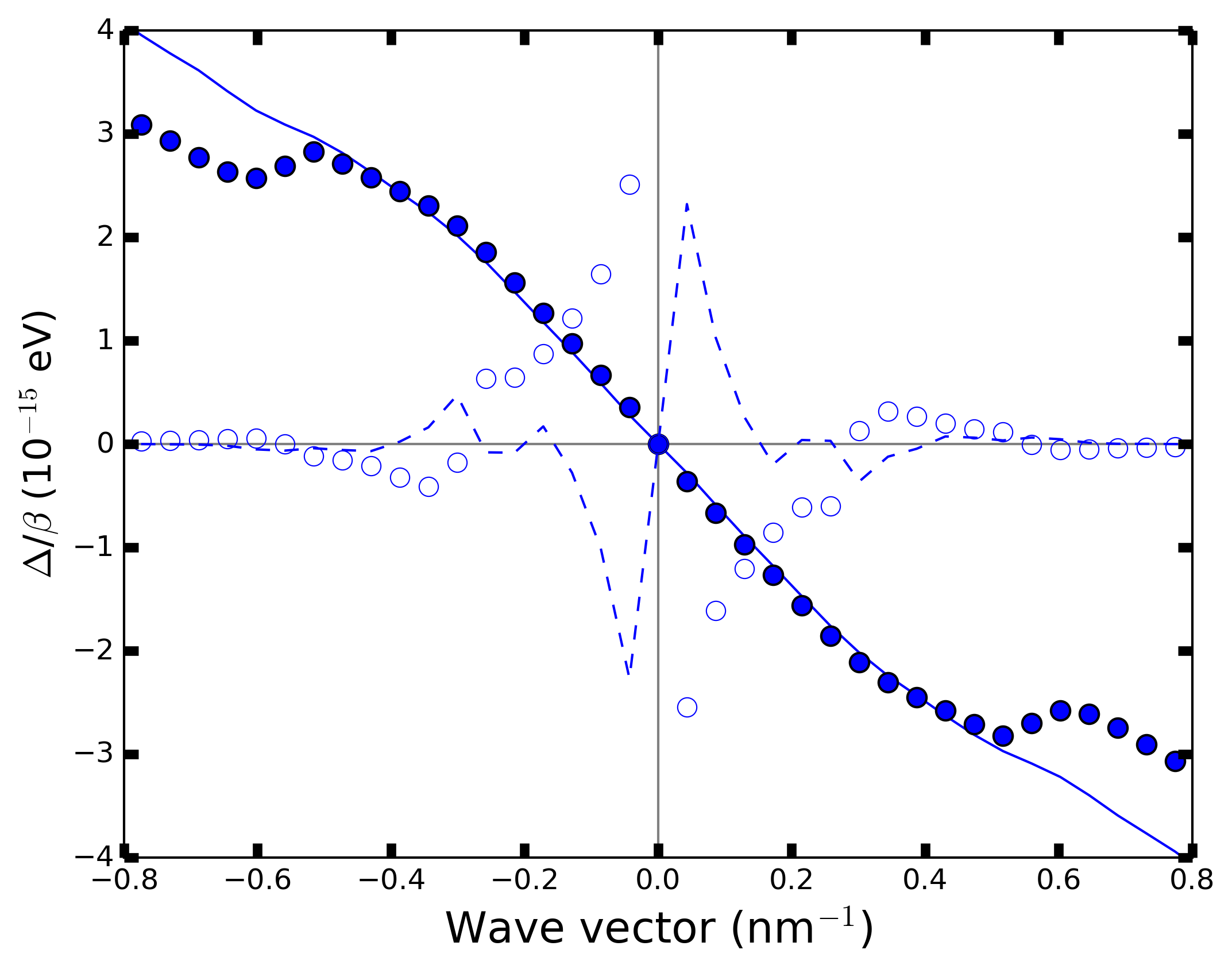}
        \caption{}
    \end{subfigure}\hfill 
    \begin{subfigure}{0.5\textwidth}
        \centering
        \includegraphics[width=0.9\textwidth]{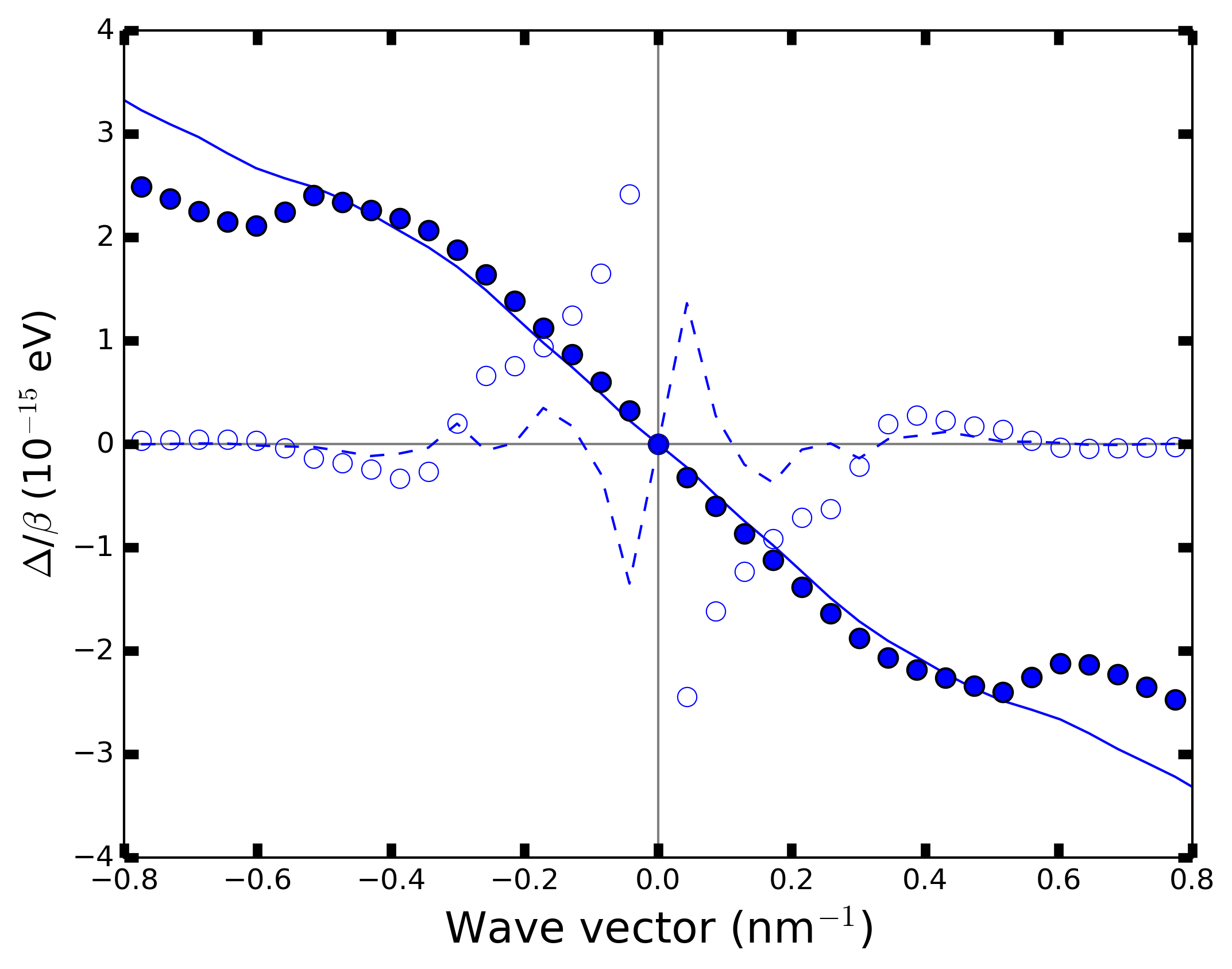}
        \caption{}
    \end{subfigure}
    \vspace{1em} 
    \begin{subfigure}{0.5\textwidth}
        \centering
        \includegraphics[width=0.9\textwidth]{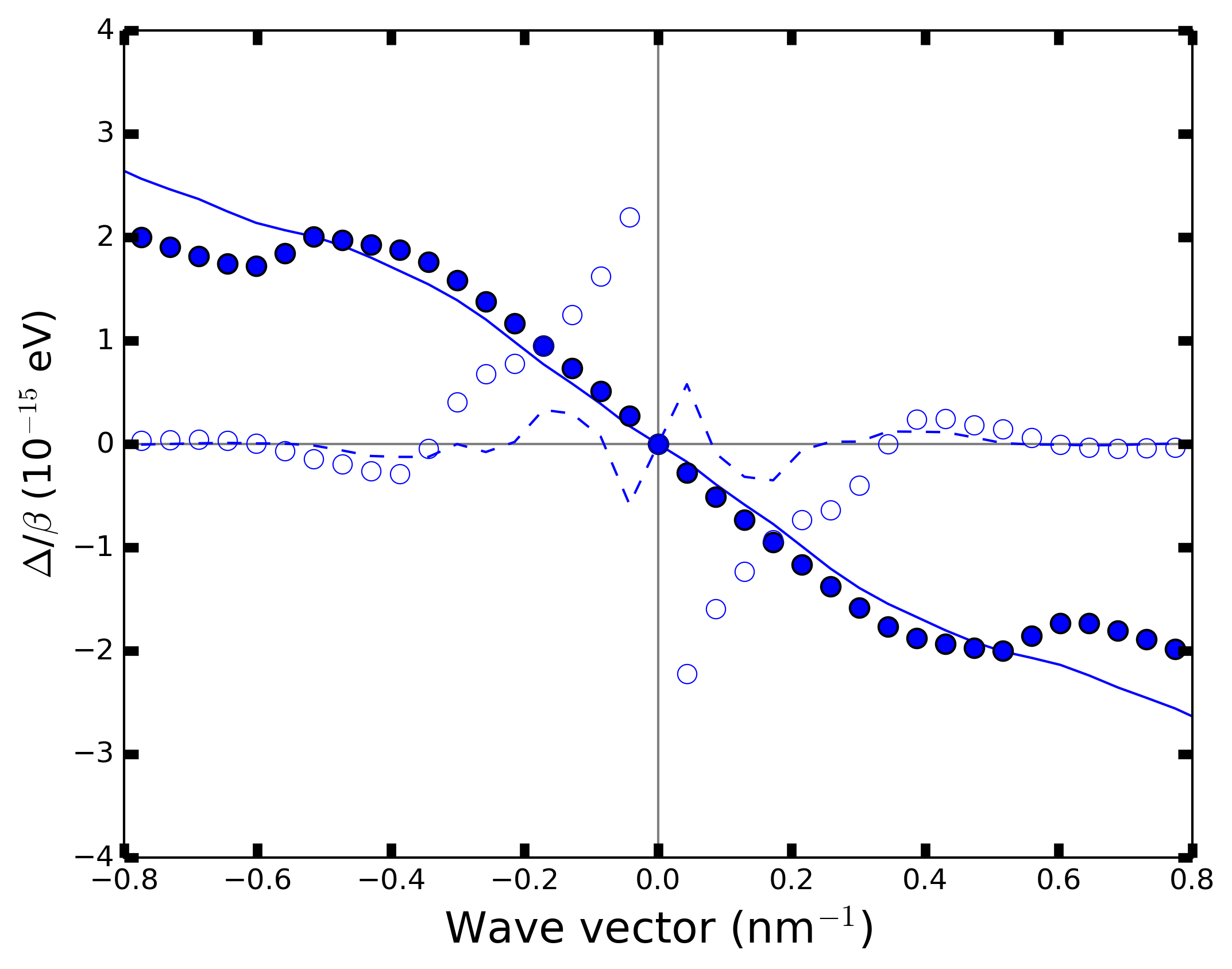}
        \caption{}
    \end{subfigure}\hfill
    \begin{subfigure}{0.5\textwidth}
        \centering
        \includegraphics[width=0.9\textwidth]{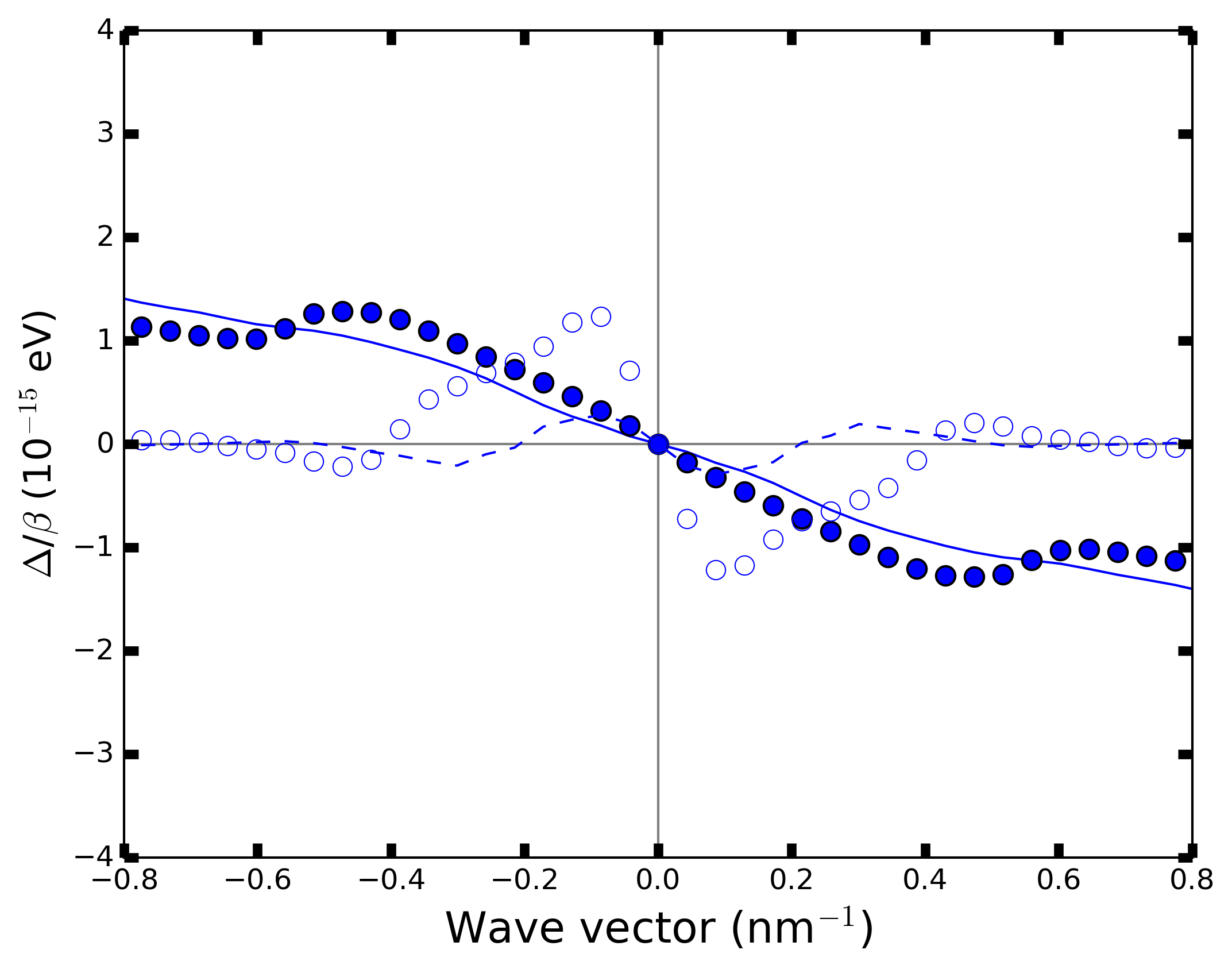}
        \caption{}
    \end{subfigure}
    \caption{\parbox{\textwidth}{Normalized deviation function $\Delta$ vs. magnitude of electron wavevector, measured from \textbf{K} point and along the (\textbf{K}-$k_x$)--\textbf{K}--(\textbf{K}+$k_x$) direction of the Brillouin zone, for electron concentrations (\textbf{a}) $2\times10^{11}$, (\textbf{b}) $5\times10^{11}$, (\textbf{c}) $7.5\times10^{11}$ and (\textbf{d}) $1.7\times10^{12}$ cm$^{-2}$ respectively at $200$ K. In all plots, lines represent the valence band and disks represent the conduction band. Solid lines and symbols are for the Coulomb enabled theory, whereas dashed lines and empty symbols are for the Coulomb-free theory.}}
    \label{dev_200K}
\end{figure*}

\begin{figure*}[ht!]
    \centering
    \begin{subfigure}{0.5\textwidth} 
        \centering
        \includegraphics[width=0.9\textwidth]{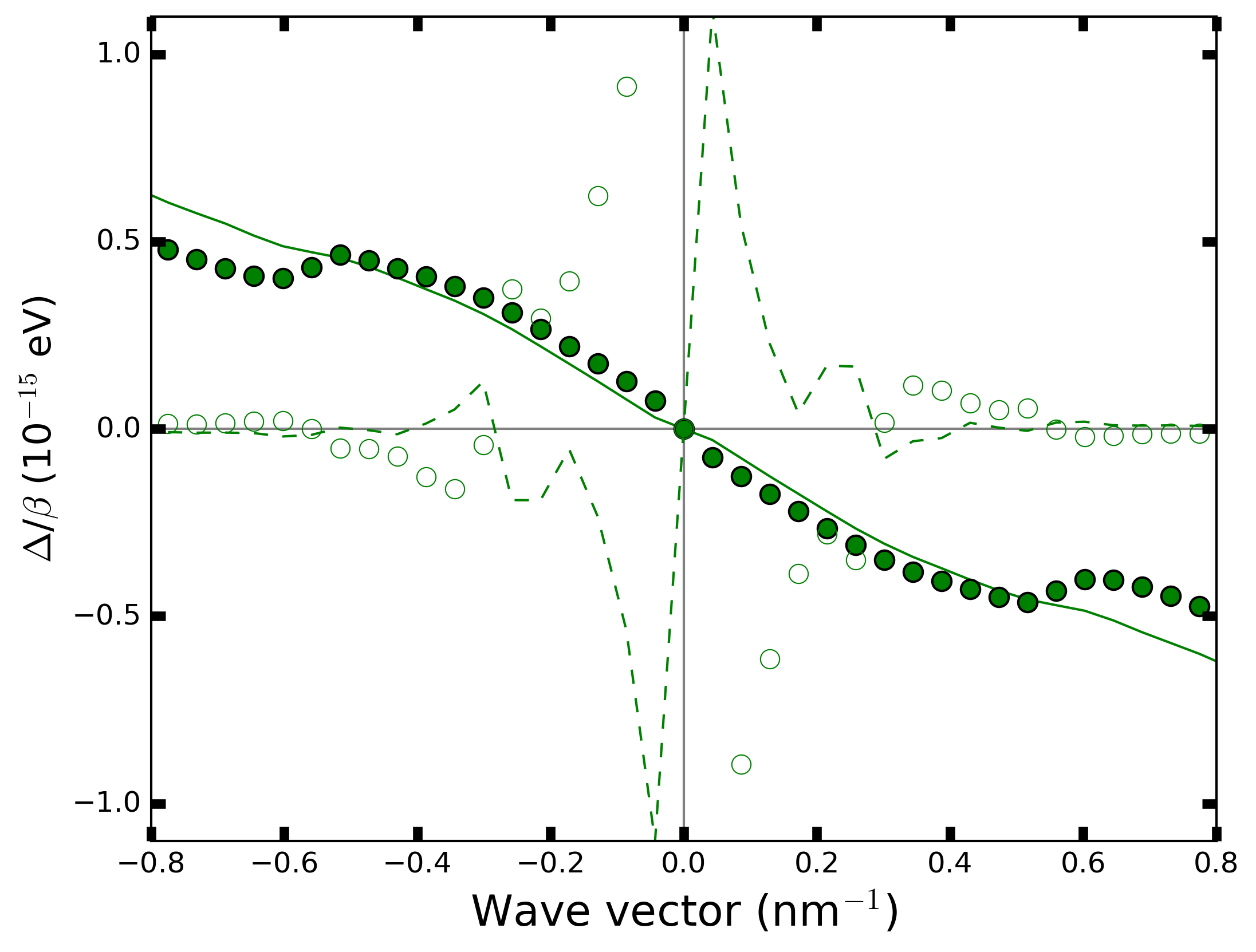}
        \caption{}
    \end{subfigure}\hfill 
    \begin{subfigure}{0.5\textwidth}
        \centering
        \includegraphics[width=0.9\textwidth]{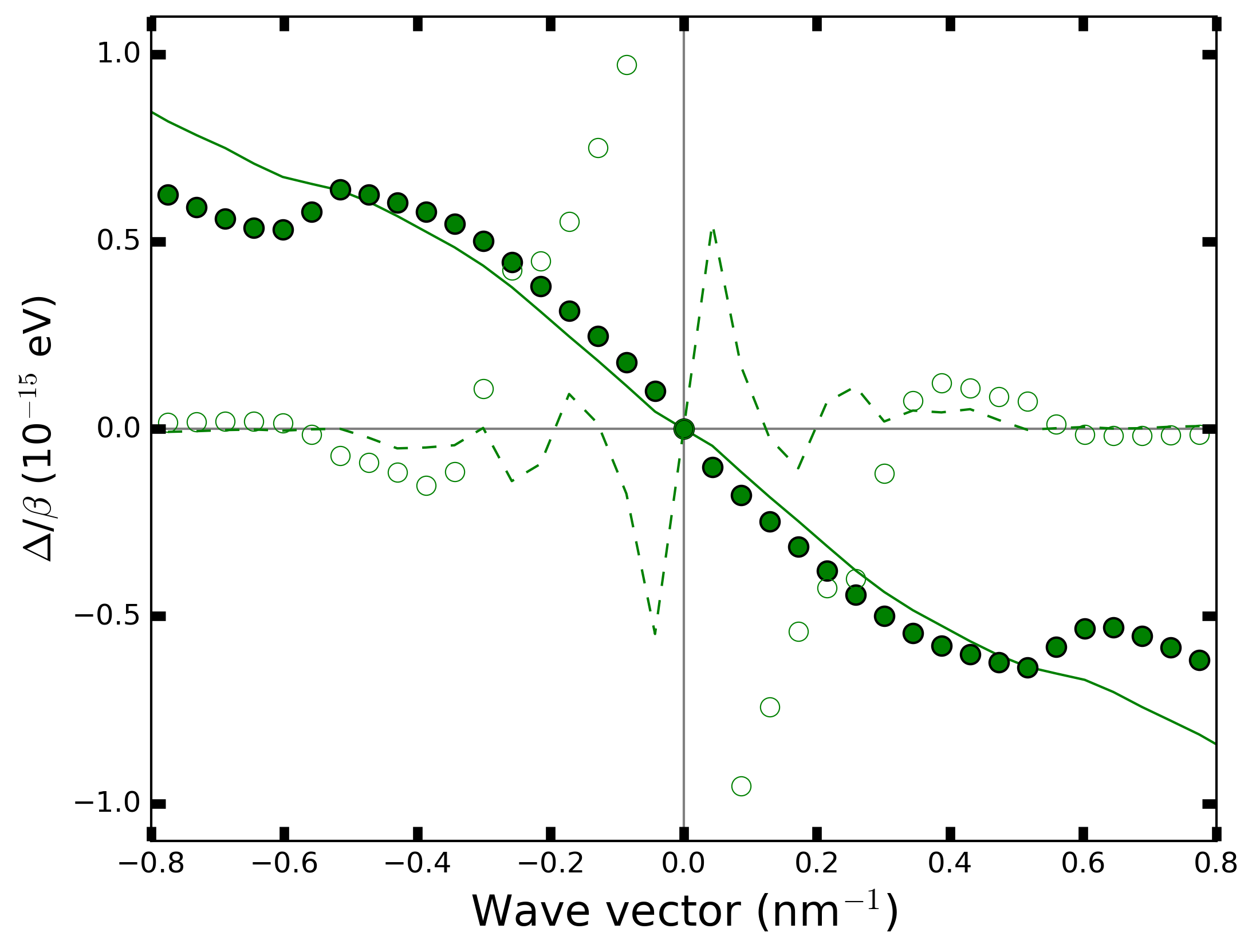}
        \caption{}
    \end{subfigure}
    \vspace{1em} 
    \begin{subfigure}{0.5\textwidth}
        \centering
        \includegraphics[width=0.9\textwidth]{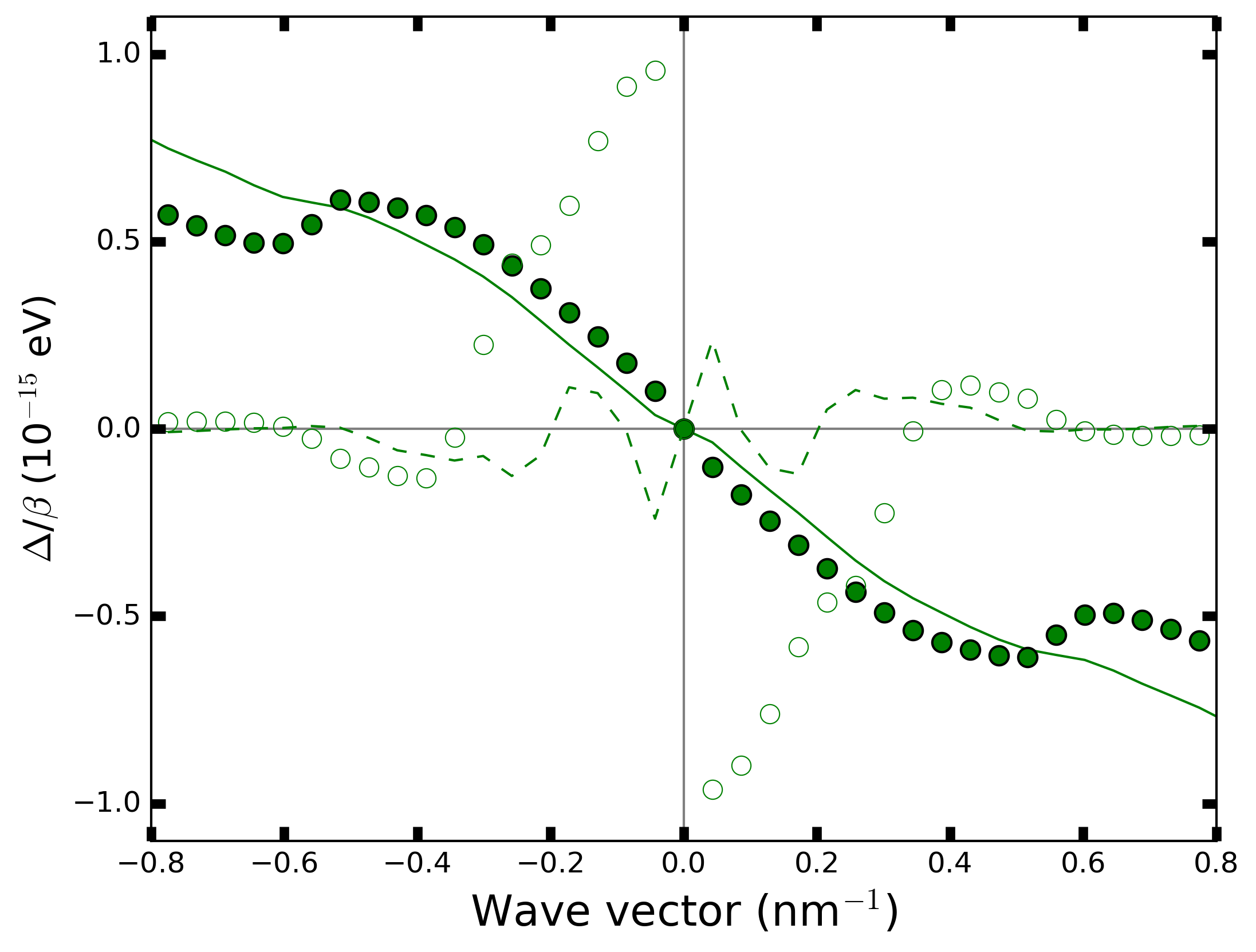}
        \caption{}
    \end{subfigure}\hfill
    \begin{subfigure}{0.5\textwidth}
        \centering
        \includegraphics[width=0.9\textwidth]{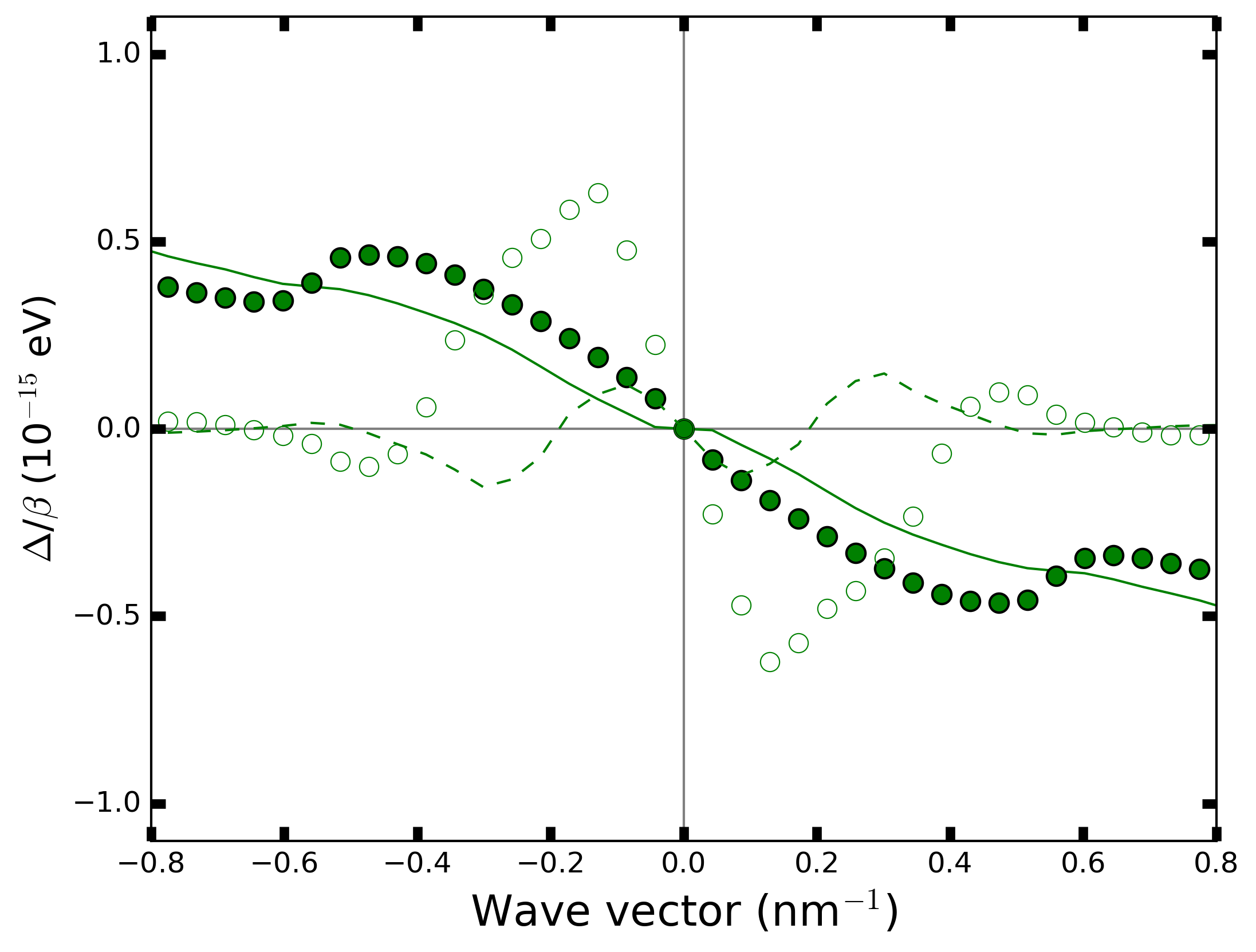}
        \caption{}
    \end{subfigure}
    \caption{\parbox{\textwidth}{Normalized deviation function $\Delta$ vs. magnitude of electron wavevector, measured from \textbf{K} point and along the (\textbf{K}-$k_x$)--\textbf{K}--(\textbf{K}+$k_x$) direction of the Brillouin zone, for electron concentrations (\textbf{a}) $2\times10^{11}$, (\textbf{b}) $5\times10^{11}$, (\textbf{c}) $7.5\times10^{11}$ and (\textbf{d}) $1.7\times10^{12}$ cm$^{-2}$ respectively at $300$ K. In all plots, lines represent the valence band and disks represent the conduction band. Solid lines and symbols are for the Coulomb enabled theory, whereas dashed lines and empty symbols are for the Coulomb-free theory.}}
    \label{dev_300K}
\end{figure*}

\begin{figure*}[ht!]
    \centering
    \begin{subfigure}{0.5\textwidth} 
        \centering
        \includegraphics[width=0.9\textwidth]{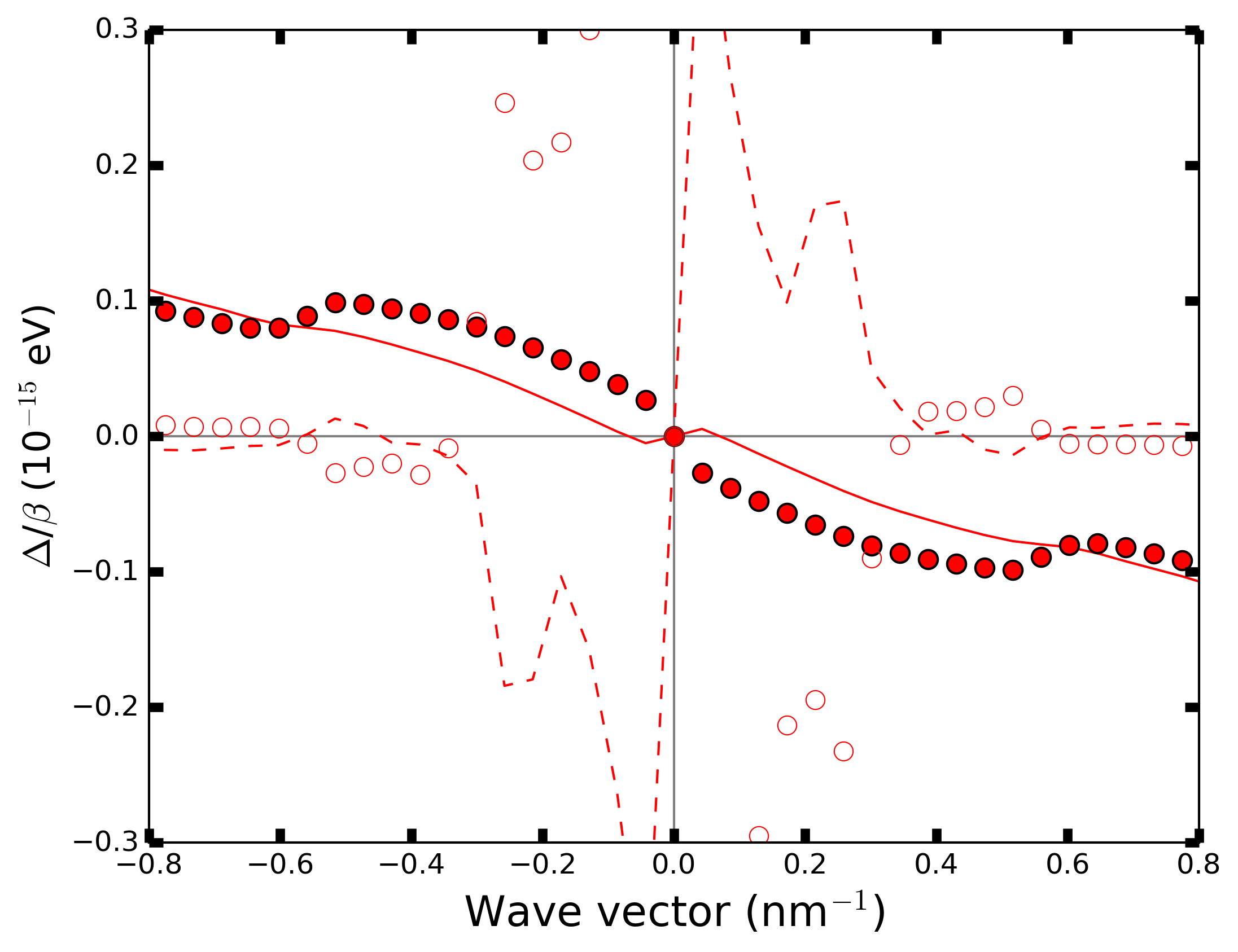}
        \caption{}
    \end{subfigure}\hfill 
    \begin{subfigure}{0.5\textwidth}
        \centering
        \includegraphics[width=0.9\textwidth]{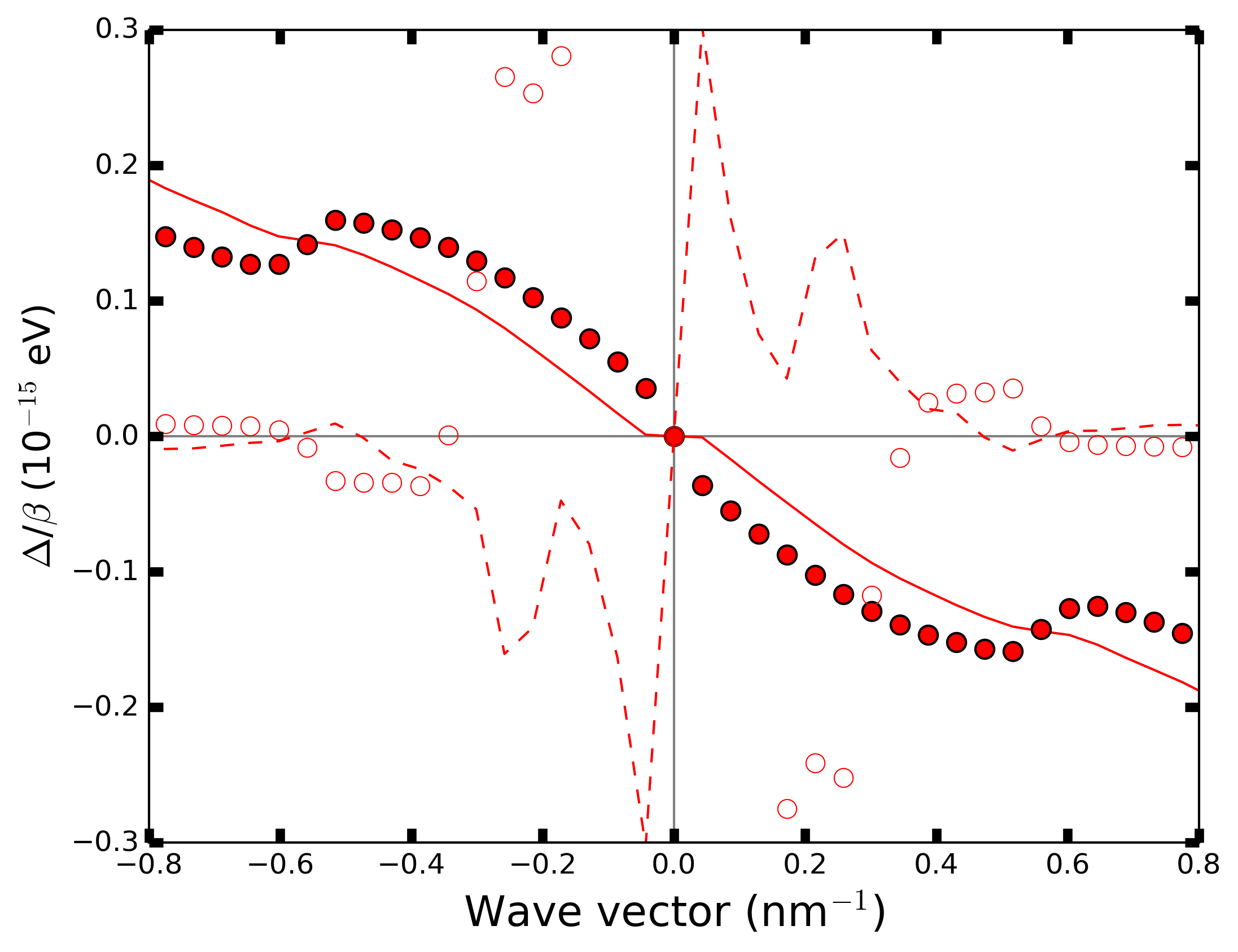}
        \caption{}
    \end{subfigure}
    \vspace{1em} 
    \begin{subfigure}{0.5\textwidth}
        \centering
        \includegraphics[width=0.9\textwidth]{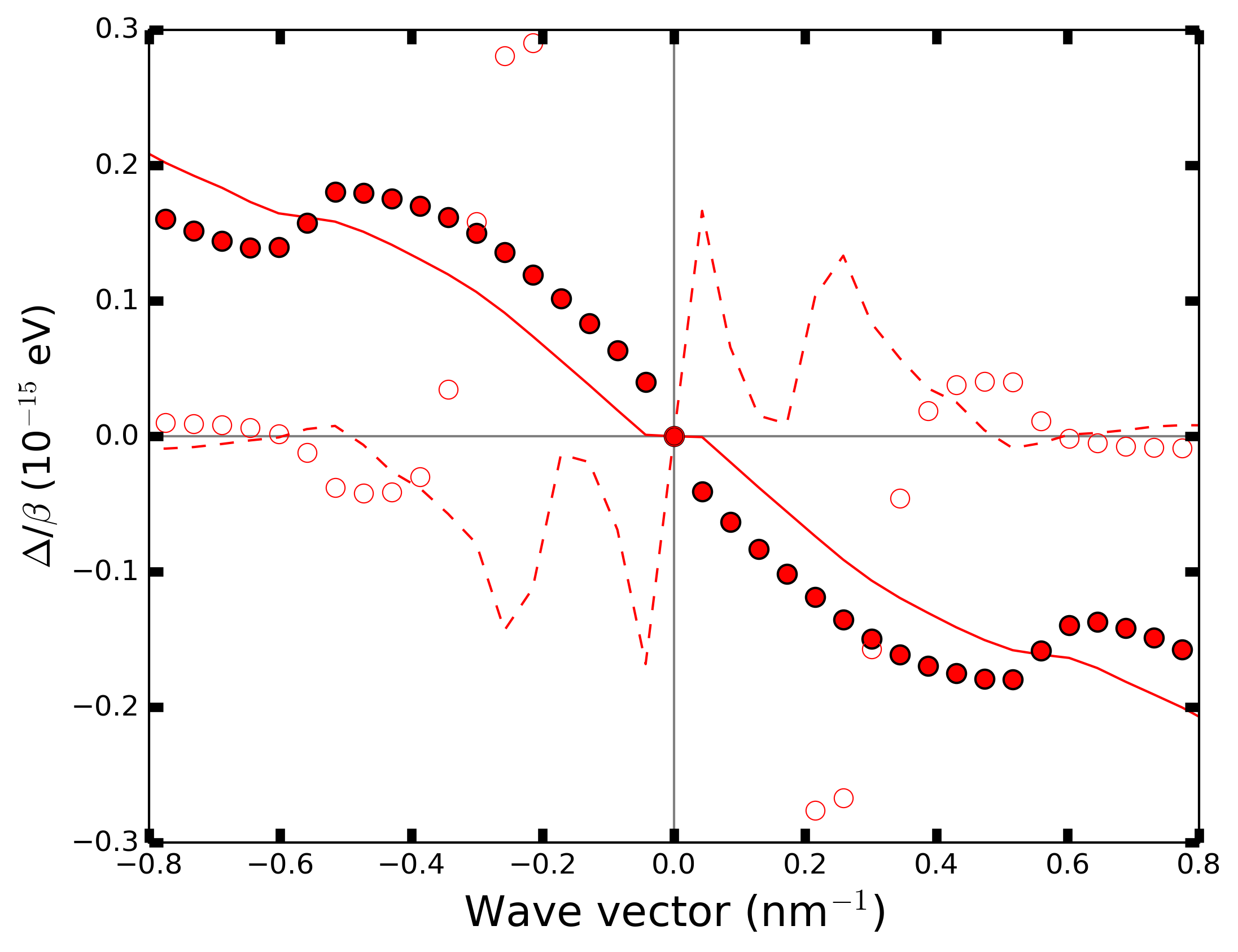}
        \caption{}
    \end{subfigure}\hfill
    \begin{subfigure}{0.5\textwidth}
        \centering
        \includegraphics[width=0.9\textwidth]{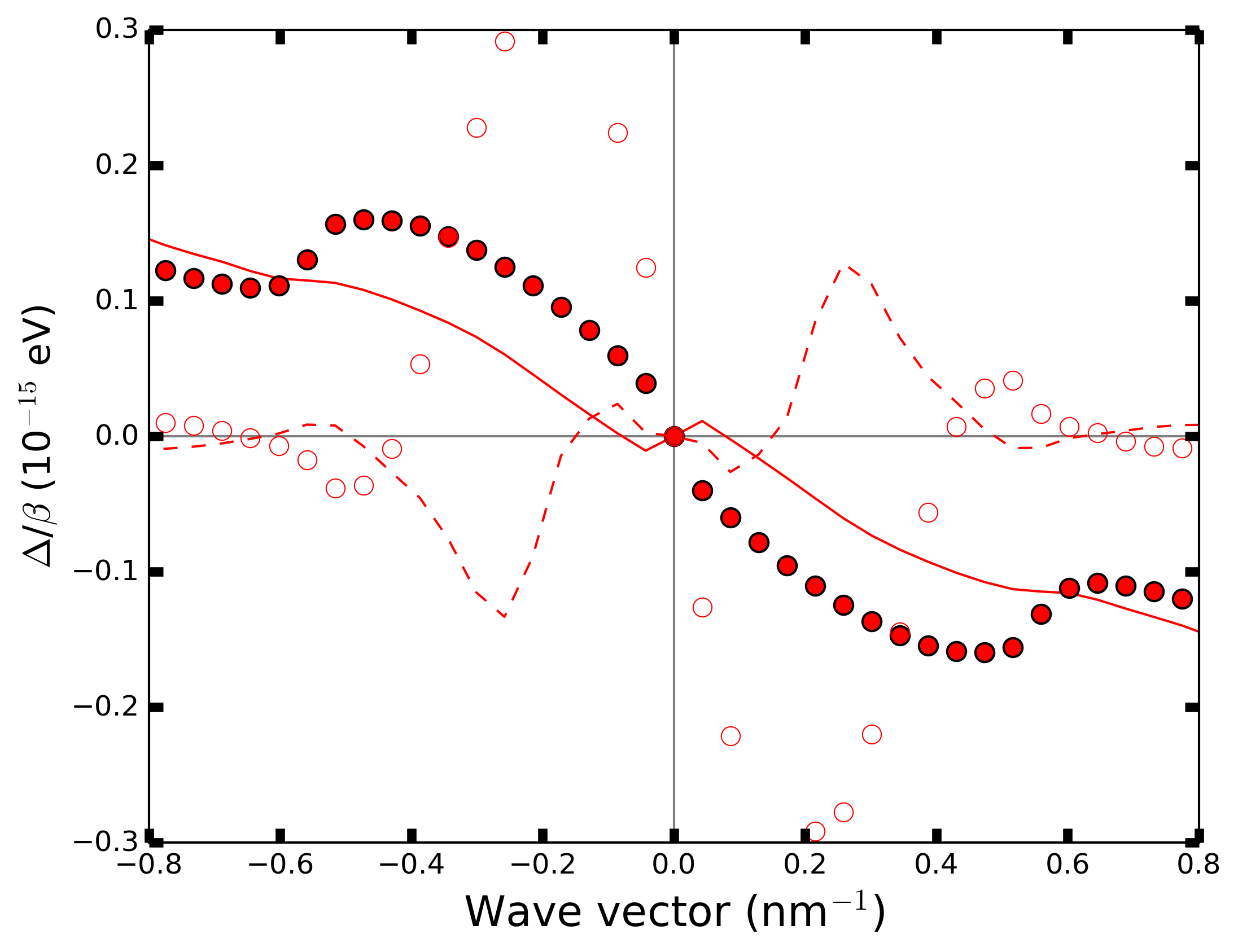}
        \caption{}
    \end{subfigure}
    \caption{\parbox{\textwidth}{Normalized deviation function $\Delta$ vs. magnitude of electron wavevector, measured from \textbf{K} point and along the (\textbf{K}-$k_x$)--\textbf{K}--(\textbf{K}+$k_x$) direction of the Brillouin zone, for electron concentrations (\textbf{a}) $2\times10^{11}$, (\textbf{b}) $4\times10^{11}$, (\textbf{c}) $7\times10^{11}$ and (\textbf{d}) $1.7\times10^{12}$ cm$^{-2}$ respectively at $400$ K. In all plots, lines represent the valence band and disks represent the conduction band. Solid lines and symbols are for the Coulomb enabled theory, whereas dashed lines and empty symbols are for the Coulomb-free theory.}}
    \label{dev_400K}
\end{figure*}

In Fig \ref{dev_200K}, \ref{dev_300K}, and \ref{dev_400K}, we present plots of normalized deviation function $\Delta/\beta$ for different concentrations of electron doped graphene at $200$, $300$, and $400$ K, respectively. As an analogue of the hole doped case presented in the the main text, here we also observe the individual and bifluid hydrodynamics when the Coulomb interactions are turned on (solid disks and solid lines). We notice that, as we go from the low doped (panel a, closest to charge neutrality) toward high doped (panel d, farthest from charge neutrality), we retain robust hydrodynamics in the electrons but begin to lose it in the holes. Hydrodynamic behavior is the weakest at the highest temperature.

To quantify the linearity of the normalized deviation function with respect to the magnitude of the wavevector, we need to see the square of Pearson correlation coefficients ($r^2$) which ranges from $0$ (non linear) to $1$ (perfectly linear). In table \ref{tab:person}, we have provided with $r^2$ for different temperature and charge carrier concentration points, for majority (minority) charge carriers when we include Coulomb interactions. We can see from the table that hydrodynamic behavior is a low temperature phenomenon. The concentration has almost negligible effect on the emergence of the individual hydrodynamic state of the charge carriers.

\begin{table}[h!]
    \centering
    \begin{tabular}{|c|c|c|c|}
    \hline
\diagbox{$n$ ($10^{12}$ cm$^{-2}$)}{$T$ (K)} & 200 & 300 & 400 \\ \hline
        0.2 &  0.9988 (0.9966) &0.9988 (0.9933) & 0.9744 (0.9759)\\
        0.3 & 0.9988 (0.9964) &0.999 (0.994) & 0.9924 (0.9835)\\
        0.4& 0.9989 (0.9965) & 0.999 (0.9939)& 0.9948 (0.9853)\\
        0.5& 0.9990 (0.9966) & 0.999 (0.9938)& 0.9955 (0.9858) \\
        0.74& 0.9992 (0.9973) & 0.9989 (0.9937)& 0.9952 (0.986)\\
        1.3& 0.9991 (0.9981) & 0.9972 (0.9938)&0.9884 (0.9846)\\
        1.7& 0.9984 (0.9981) & 0.9943 (0.9936)& 0.9771 (0.9837)\\
        \hline
    \end{tabular}
    \caption{Pearson correlation coefficient of linear regression for normalized deviation function versus magnitude of majority (minority) charge carrier wave vector for different temperatures ($T$) and charge carrier concentrations ($n$).}
    \label{tab:person}
\end{table}


\section{Correction algorithms based on Kelvin-Onsager reciprocal relations}

The Kelvin-Onsager relation (KOR) is given by the following expression:
\begin{align}
    \sigma S  = \dfrac{\alpha_{\text{total}}}{T}\label{K-O3}, 
\end{align}
where $\sigma S$ is the thermoelectric transport coefficient related to the electric response to an applied temperature gradient. This is related to the Seebeck effect. And $\alpha_{\text{total}} = \alpha_{\text{el}} + \alpha_{\text{ph}}$ is related to the reciprocal, Peltier thermoelectric effect which, as it turns out, cleanly splits into purely electronic and phononic components. When phonon drag is ignored, as is the case in this work, $\alpha_{\text{ph}} = 0$.

We solve the BTE, strictly enforcing the KOR. To do so, we introduce a correction via the matrix $M_{\mathbf{I}}$ for the linear response function $\mathbf{I}$ of the electronic subsystem with respect to the temperature gradient, such that $\mathbf{I}_{m\mathbf{k}}' = M_{\mathbf{I}}\mathbf{I}_{m\mathbf{k}}$ and $\sigma S[\mathbf{I'}]$ satisfies Eq. \ref{K-O3}. Here $m$ and $\mathbf{k}$ stand for the band and wavevector, respectively. 

The general form of the dependence of a transport coefficient $a$ on the corresponding linear response function $\mathbf{Y}$ is as follows:

\begin{align}\label{Gen_tran_coeff}
     a[\mathbf{Y}] = \sum_{m \mathbf{k}}g(m,\mathbf{k}) \mathbf{v}_{m \mathbf{k}}\otimes \mathbf{Y}_{m\mathbf{k}}, 
\end{align}
where $\mathbf{v}_{m\mathbf{k}}$ is the group velocity, $\mathbf{Y}_{m\mathbf{k}}$ is the band- and wavevector-resolved linear response function, and the scalar function $g(m,\mathbf{k})$ absorbs other coefficients and the occupation-dependent part of the expressions. 

Using Eq. \ref{Gen_tran_coeff}, we can derive how $a$ changes under the transformation   $\mathbf{Y}_{m\mathbf{k}}' = M_{\mathbf{Y}} \mathbf{Y}_{m\mathbf{k}}$:
\begin{align}
    & (\mathbf{v}\otimes \mathbf{Y'})_{\alpha\beta} = (\mathbf{v}\otimes M_\mathbf{Y}\mathbf{Y})_{\alpha\beta} = \mathbf{v}_\alpha \sum_{k} (M_{\mathbf{Y}})_{\beta\gamma}\mathbf{Y}_\gamma = \mathbf{v}_\alpha \sum_{\gamma}\mathbf{Y}_\gamma (M^T_{\mathbf{Y}})_{\gamma\beta} = ([\mathbf{v}\otimes \mathbf{Y}]M_\mathbf{Y}^T)_{\alpha\beta},
\end{align}
where $\alpha$, $\beta$, and $\gamma$ are Cartesian indices.

Consequently,
\begin{align}
    & \sigma S[\mathbf{I}'] = \sigma S[\mathbf{I}]M_{\mathbf{I}}^T
\end{align}

There are two important constraints for the corrections. The first one is that the correction must preserve the symmetry of the transport coefficients, which corresponds to the crystallographic point group of the given crystal. The second one is dictated by the second law of thermodynamics, which states that the entropy production should be non‑negative:
\begin{align}\label{Entropy}
    T\dot{\delta S} = \sum_j \mathbf{X}_j^T \mathbf{J}_j  \geq 0,
\end{align}
where $\mathbf{X}_j$ denotes a thermodynamic force and $\mathbf{J}_j$ the associated flux of various tensorial order \cite{2nd_law}.

In the linear response regime, fluxes are linearly proportional to the forces, that is $\mathbf{J}_j = \sum_k A_{jk}\mathbf{X}_k$, where each $A_{jk}$ is a matrix (Cartesian basis) of coefficients. Substituting this into Eq. \ref{Entropy}, we get
\begin{align}\label{Entropy2}
    T\dot{\delta S} = \sum_{j,k} \mathbf{X}_j^T A_{jk}\mathbf{X}_k \geq 0, \forall \mathbf{X}_j
\end{align}

This means that the block matrix $A$ is  positive semi-definite. In the given case $\mathbf{X}_1 = \mathbf{E}$ and $\mathbf{X}_2 = -\nabla T/T$ and the matrix of coefficients is
\begin{align}\label{L_mat2}
   A = \begin{bmatrix}
\sigma & T\sigma S(\mathbf{I'})  \\
 \alpha_{\text{el}} + \alpha_{\text{ph}}  & T(\kappa_{\text{el}}^{0}(\mathbf{I'}) + \kappa_{ph})
\end{bmatrix}
\end{align}

As a result, our correction algorithm corresponds to solving the matrix equation $\sigma S[\mathbf{I}]M^T_{\mathbf{I}} = \alpha T^{-1}$ with respect to $M_{\mathbf{I}}$, which also satisfies the constraints dictated by the second law of thermodynamics and the symmetry of the transport coefficients.

It is worth noting that both the electrons and the phonon subsystems contribute to the entropy production. As such, the thermal conductivities of both should be included. This, however, does not necessarily mean that the correction will be different if one of the subsystems is ignored. This is because Eq. \ref{Entropy2} is an inequality condition. In our current case, we have verified that regardless of whether $\kappa_{\text{ph}}$ is set to $0$ or a large value of $3000$ Wm$^{-1}$K$^{-1}$, we get identical corrections.

\bibliography{supbib}